\documentstyle[floats,prd,aps,preprint,epsf]{revtex}

\draft
\begin{document}
%@twocolumnfalse\endcsname

\newcommand{\beq}{\begin{equation}}
\newcommand{\eeq}{\end{equation}}
\newcommand{\beqn}{\begin{eqnarray}}
\newcommand{\eeqn}{\end{eqnarray}}
\newcommand{\pa}{\partial}
\newcommand{\vp}{\varphi}
\newcommand{\varep}{\varepsilon}
\def\zero{\hbox{$_{(0)}$}}
\def\bL{\hbox{$\,{\cal L}\!\!\!$--}}
\def\bI{\hbox{$\,I\!\!\!$--}}

\begin{center}
{\large\bf{Gravitational waves from axisymmetric rotating stellar 
core collapse to a neutron star in full general relativity 
}}
~\\
~\\
Masaru Shibata and Yu-ichirou Sekiguchi\\
{\em Graduate School of Arts and 
Sciences, University of Tokyo, Tokyo, 153-8902, Japan}\\
\end{center}
\begin{abstract}
Axisymmetric numerical simulations of rotating 
stellar core collapse to a neutron star are performed
in the framework of full general relativity. 
The so-called Cartoon method, in which the Einstein 
field equations are solved in the Cartesian coordinates 
and the axisymmetric condition is imposed around the $y=0$ plane,
is adopted. 
The hydrodynamic equations are solved in the cylindrical coordinates 
(on the $y=0$ plane in the Cartesian coordinates) 
using a high-resolution shock-capturing scheme 
with the maximum grid size $(2500,2500)$. 
A parametric equation of state is adopted to model 
collapsing stellar cores and neutron stars following Dimmelmeier et al. 
It is found that the evolution of central density during the collapse, 
bounce, and formation of protoneutron stars agree 
well with those in the work of Dimmelmeier et al. 
in which an approximate general relativistic formulation is adopted. 
This indicates that such approximation is appropriate 
for following axisymmetric stellar core collapses and subsequent
formation of protoneutron stars. 
Gravitational waves are computed using a quadrupole formula. 
It is found that the waveforms are qualitatively in good agreement 
with those by Dimmelmeier et al. However, quantitatively,
two waveforms do not agree well.
Possible reasons for the disagreement are discussed. 
\end{abstract}
%\vskip2pc]

\section{Introduction}

Rotating stellar core collapse is among the most 
promising sources of gravitational waves. 
To date, there has been no systematic work for computation 
of the collapse to a neutron star and of emitted gravitational waves 
in full general relativity (but see \cite{Siebel}). 
Gravitational waves associated with the formation of rotating 
neutron stars have been widely computed in the Newtonian gravity 
\cite{Newton,Newton1,Newton2,Newton3,Newton4,Newton45,Newton5,Newton6} 
or in an approximate general relativistic gravity \cite{HD} 
using the so-called conformal flatness approximation 
(or Isenberg-Wilson-Mathews approximation \cite{IWM}). 
As indicated in \cite{HD}, the general relativistic effects modify 
the dynamics of the collapse and corresponding gravitational waveforms
significantly. This implies that simulation in full general relativity 
is the best approach for accurate computation of gravitational waves. 

During the stellar core collapse to a neutron star, the characteristic 
radius changes from initial stellar core radius $\sim 2000$ km to neutron 
star radius $\sim 10$ km. 
Adopting a uniform and fixed grid with a grid spacing as 
$\sim 1$ km, the required grid number for the simulation is 
more than 2000 for one direction. 
With current computational resources, it is very difficult to 
take such a huge number of grid points in three-dimensional simulations. 
If the progenitor of the neutron star is not very rapidly rotating, 
nonaxisymmetric instabilities will not set in and 
the collapse will proceed in an axisymmetric manner. 
By restricting our attention to axisymmetric spacetimes, 
the grid resolution can be improved significantly for a given
computational resource. Thus, as a first step, it is better to perform 
axisymmetric simulations than to do nonaxisymmetric ones for a well-resolved
and convergent computation of the collapse, the bounce,
and corresponding gravitational waveforms, 
focusing only on the moderately rapid rotation case. 

In this paper, we study gravitational waves 
from rotating stellar core collapses to a neutron star assuming 
the axial symmetry. Dynamics of the 
collapse is followed by fully general relativistic simulation. 
Gravitational waves are approximately computed using a 
quadrupole formula adopted and tested in \cite{SS}. 
Necessity of adopting quadrupole formulas arises from the fact 
that the amplitude of gravitational waves is too small 
($< 10^{-5}$ in the local wave zone) to accurately extract the
waveforms from raw data sets of metric. Although 
exact gravitational waveforms cannot be computed,
the quadrupole formula is a useful tool for
approximate computation of
gravitational waves associated with matter motion such as
oscillations of neutron stars as indicated in \cite{SS}. 

Recently, gravitational waves from axisymmetric
rotating stellar core collapses 
have been extensively computed in a relativistic manner 
by Dimmelmeier et al. \cite{HD}. As mentioned above, they
determine the gravitational fields adopting 
an approximate formulation of the Einstein equation. 
The approximation is likely to be 
applicable to a moderately relativistic and stationary spacetime such as
that for a rapidly rotating neutron star \cite{CST96}. 
However, no one has clarified whether this is the case 
for dynamical spacetimes. 
To confirm that their treatment is indeed appropriate, it is necessary
to compare their solutions with fully general relativistic ones for
a calibration. One of the purposes in this paper is to examine whether
the numerical solution for the stellar core collapse computed
in \cite{HD} is a well approximated one for a fully general
relativistic solution. 

In \cite{HD}, gravitational waveforms were computed in terms of a
quadrupole formula. 
%%which is a Newtonian-based one and different from
%%that adopted in \cite{SS} and in this paper.
In general relativity, there is no unique definition of the quadrupole 
moment, nor is the quadrupole formula, for axisymmetric
dynamical spacetimes.
Accuracy of gravitational waveforms depends on the choice of
the quadrupole formula and the gauge conditions. 
Thus, to know how accurately the approximate gravitational
waveforms can be computed by the chosen quadrupole formula,
a calibration is required by comparing the resulting waveforms
with those computed from the metric
as we did in \cite{SS}. Unfortunately, such calibration is not
possible in \cite{HD} since they adopted an approximate general
relativistic formulation for the gravitational field in which the
metric does not carry any information of gravitational waves.
Consequently, it is not clear whether the quadrupole formula
they adopted can actually yield accurate approximate gravitational waveforms
and how large the magnitude of the error is.
On the other hand, in the previous paper \cite{SS}, 
we did such a calibration for a quadrupole
formula which is different from that in \cite{HD}, and 
showed it possible to compute gravitational waves 
from oscillating and rapidly rotating neutron stars of high values of 
compactness fairly accurately, 
besides possible systematic errors for the amplitude
due to neglecting post-Newtonian corrections. 
Computing gravitational waveforms by the calibrated quadrupole formula
and comparing the results with the previous ones, 
we estimate how accurate the waveforms computed in \cite{HD}.

This paper is organized as follows. 
In Sec. II, our numerical implementations for 
general relativistic simulation in axial symmetry are briefly reviewed. 
In Sec. III, the initial condition and computational setting 
are described. Sec. IV presents the numerical results. 
Sec. V is devoted to a summary. Throughout this paper, we adopt the
geometrical units in which $G=c=1$ where $G$ and $c$ are the 
gravitational constant and the speed of light, respectively. 

\section{Numerical implementation}

\subsection{Summary of formulation}

We perform fully general relativistic simulations for rotating 
stellar core collapse in axial symmetry using the same formulation as 
in~\cite{S2002}, to which the reader may refer for details and basic 
equations. The fundamental variables for the hydrodynamics are 
\beqn 
\rho &&:{\rm rest~ mass~ density},\nonumber \\
\varep &&: {\rm specific~ internal~ energy}, \nonumber \\
P &&:{\rm pressure}, \nonumber \\
u^{\mu} &&: {\rm four~ velocity}, \nonumber \\
v^i &&={dx^i \over dt}={u^i \over u^t},
\eeqn
where subscripts $i, j, k, \cdots$ denote $x, y$ and $z$, and 
$\mu$ the spacetime components. 
As the variables to be evolved in the numerical simulations, 
we define a weighted density $\rho_*(=\rho \alpha u^t e^{6\phi})$ 
and a weighted four-velocity $\hat u_i [= (1+\varepsilon+P/\rho)u_i]$. 
From these variables, the total baryon rest-mass and angular momentum 
of system, which are conserved quantities in axisymmetric
spacetimes, can be defined as 
\beqn
M_*&=&\int d^3 x \rho_*, \\
J  &=&\int d^3 x \rho_*\hat u_{\varphi}. 
\eeqn
General relativistic hydrodynamic equations are solved using 
a so-called high-resolution shock-capturing scheme \cite{Font,S2002} 
on the $y=0$ plane with the cylindrical coordinates $(x, z)$
(in the Cartesian coordinates with $y=0$). 

The fundamental variables for geometry are 
\beqn
\alpha &&: {\rm lapse~function}, \nonumber \\
\beta^k &&: {\rm shift~vector}, \nonumber \\
\gamma_{ij} &&:{\rm metric~in~three~dimensional~spatial~hypersurface},
\nonumber \\ 
\gamma &&=e^{12\phi}={\rm det}(\gamma_{ij}), \nonumber \\
\tilde \gamma_{ij}&&=e^{-4\phi}\gamma_{ij}, \nonumber \\
K_{ij} &&:{\rm extrinsic~curvature}. 
\eeqn
We evolve $\tilde \gamma_{ij}$, $\phi$, 
$\tilde A_{ij} \equiv e^{-4\phi}(K_{ij}-\gamma_{ij} K_k^{~k})$,
and trace of the extrinsic curvature $K_k^{~k}$ 
together with three auxiliary functions
$F_i\equiv \delta^{jk}\pa_{j} \tilde \gamma_{ik}$ with an
unconstrained free evolution code
as in \cite{shibata,SBS,bina,SN,shiba2000,S2002}. 

The Einstein equations are solved in the Cartesian coordinates. 
To impose axisymmetric boundary conditions, the Cartoon method
is used \cite{alcu}: Assuming the reflection symmetry with 
respect to the equatorial plane, simulations are performed 
using a fixed uniform grid with the grid 
size $N \times 3 \times N$ in $(x, y, z)$ which covers 
a computational domain as 
$0 \leq x \leq L$, $0 \leq z \leq L$, and $-\Delta \leq y \leq \Delta$. 
Here, $N$ and $L$ are constants and $\Delta = L/N$. 
In the Cartoon method, the axisymmetric boundary conditions are
imposed at $y= \pm \Delta$. 

As the time slice, we impose 
an ``approximate'' maximal slicing condition in which 
$K_k^{~k} \approx 0$ is required \cite{shibata}. 
As the spatial gauge, 
we adopt a dynamical gauge condition \cite{S03} in which
the equation for the shift vector is written as 
\beq
\pa_t \beta^k = \tilde \gamma^{kl} (F_l +\Delta t \pa_t F_l),
\label{dyn}
\eeq
where $\Delta t$ denotes a timestep in numerical computation. 

During the numerical simulations, 
violations of the Hamiltonian constraint and conservation of
mass and angular momentum are monitored as code checks.
Numerical results for several test
calculations, including stability and collapse of nonrotating
and rotating neutron stars, have been described in~\cite{S2002}. 

An outgoing wave boundary condition 
for $F_i$, $h_{ij}(=\tilde \gamma_{ij}-\delta_{ij})$, and $\tilde A_{ij}$
is imposed at the outer boundaries of the computational domain. 
The condition adopted is the same as that described in \cite{SN}. 
$K_k^{~k}$ is set to be zero at the outer boundaries. 

\subsection{Equations of state}

A parametric equation of state is adopted following M\"uller and
his collaborators \cite{Newton4,HD}. In this equation of state,
one assumes that the pressure consists of the sum of polytropic and
thermal parts as
\beq
P=P_{\rm P}+P_{\rm th}. \label{EOSII}
\eeq
The polytropic part is given by 
$P_{\rm P}=K_{\rm P}(\rho) \rho^{\Gamma(\rho)}$
where $K_{\rm P}$ and $\Gamma$ are not constants but functions of $\rho$.
In this paper, we follow \cite{HD} for the choice of 
$K_{\rm P}(\rho)$ and $\Gamma(\rho)$: 
For the density smaller than the nuclear density which is defined as 
$\rho_{\rm nuc} \equiv 2\times 10^{14}~{\rm g/cm^3}$, 
$\Gamma=\Gamma_1(=$const) is set to be $\alt 4/3$, and 
for $\rho \geq \rho_{\rm nuc}$, $\Gamma=\Gamma_2(={\rm const}) \geq 2$.
Thus,
\beqn
P_{\rm P}=
\left\{
\begin{array}{ll}
K_1 \rho^{\Gamma_1}, & \rho \leq \rho_{\rm nuc}, \\
K_2 \rho^{\Gamma_2}, & \rho \geq \rho_{\rm nuc}, \\
\end{array}
\right.\label{P12EOS}
\eeqn
where $K_1$ and $K_2$ are constants. 
Since $P_{\rm P}$ should be continuous, the relation, 
$K_2=K_1\rho_{\rm nuc}^{\Gamma_1-\Gamma_2}$, is required. 
Following \cite{Newton4,HD}, the value of $K_1$ is fixed to 
$5\times 10^{14}$ cgs. With this choice, 
the polytropic part of the equation of state for 
$\rho < \rho_{\rm nuc}$, in which the degenerate pressure of electrons is 
dominant, is approximated well. Since the specific internal energy 
should be continuous at $\rho=\rho_{\rm nuc}$, the polytropic specific 
internal energy $\varepsilon_{\rm P}$ is defined as 
\beqn
\varepsilon_{\rm P}=
\left\{
\begin{array}{ll}
\displaystyle
{K_1 \over \Gamma_1-1} \rho^{\Gamma_1-1}, & \rho \leq \rho_{\rm nuc}, \\
\displaystyle 
{K_2 \over \Gamma_2-1} \rho^{\Gamma_2-1}
+{(\Gamma_2-\Gamma_1)K_1 \rho_{\rm nuc}^{\Gamma_1-1}
\over (\Gamma_1-1)(\Gamma_2-1)},  & \rho \geq \rho_{\rm nuc}. \\
\end{array}
\right.
\eeqn
With this setting, a realistic equation of state for high-density,
cold nuclear matter is mimicked. 

The thermal part of the pressure $P_{\rm th}$ plays an important
role in the case
that shocks are generated. $P_{\rm th}$ is related to the thermal energy
density $\varepsilon_{\rm th}\equiv \varepsilon-\varepsilon_{\rm P}$ as 
\beq
P_{\rm th}=(\Gamma_{\rm th}-1)\rho \varepsilon_{\rm th}. 
\eeq
Following \cite{HD}, the value of $\Gamma_{\rm th}$, which determines
the strength of shocks, is chosen as
$1.5$ for most simulations in this paper. Extending the previous work
\cite{HD}, for a few models, 
we set $\Gamma_{\rm th}=1.35$ or 5/3 to investigate the effect
of the shock heating at  
the bounce phase and resulting gravitational waveforms. 

Simulations are initiated in the following manner: First, 
equilibrium rotating stars with $\Gamma=4/3$ polytrope are given.
Then, to induce the collapse, we slightly decrease the value of
the adiabatic index from $\Gamma=4/3$ to $\Gamma_1 < 4/3$. 
The equilibrium states are computed with the polytropic equation of state as 
\beq
P=K_0 \rho^{4/3},\label{EOS43}
\eeq
where following \cite{HD}, 
$K_0$ is set to be $5 \times 10^{14}~{\rm cm^3/s^2/gr^{1/3}}$, with which
a soft equation of state governed by the electron degenerate pressure
is approximated well \cite{ST}. Here, $K_0$ and $K_1$ are related
by $K_1=K_0\rho_0^{4/3-\Gamma_1}$ where we set $\rho_0=1~{\rm g/cm^3}$. 

\subsection{Quadrupole formula}

In the present work, gravitational waveforms are computed 
using a quadrupole formula \cite{SS}.
In quadrupole formulas, gravitational waves
at null infinity are calculated from 
\beq
h_{ij}=P_{i}^{~k} P_{j}^{~l} \biggl(
{2 \over r}{d^2\bI_{kl} \over dt^2}\biggr),\label{quadf}
\eeq
where $\bI_{ij}$ and $P_i^{~j}=\delta_{i}^{~j}-n_i n^j$ ($n^i=x^i/r$) 
denote a tracefree quadrupole moment and the projection tensor. 
From this expression, the $+$-mode of gravitational waves with 
$l=2$ in axisymmetric spacetimes is written as 
\beq
h_+^{\rm quad} = {\ddot I_{xx}(t_{\rm ret}) - \ddot I_{zz}(t_{\rm ret})
\over r}\sin^2\theta, \label{quadr}
\eeq
where $I_{ij}$ denotes a quadrupole moment, $\ddot I_{ij}$ 
its second time derivative, and $t_{\rm ret}$ a retarded time. 

In fully general relativistic and dynamical spacetimes, 
there is no unique definition for the quadrupole moment and 
nor is for $\ddot I_{ij}$. Following a previous paper \cite{SS}, 
we choose the simplest definition as 
\beq
I_{ij} = \int \rho_* x^i x^j d^3x. 
\eeq
Then, using the continuity equation of the form 
\beq
\pa_t \rho_* + \pa_i (\rho_* v^i)=0, 
\eeq
the first time derivative can be written as 
\beq
\dot I_{ij} = \int \rho_* (v^i x^j +x^i v^j)d^3x.
\eeq
To compute $\ddot I_{ij}$, the finite differencing of the numerical result 
for $\dot I_{ij}$ is carried out. 

As indicated in \cite{SS}, it is possible to compute gravitational waves 
from oscillating and rapidly rotating neutron stars of high values of 
compactness fairly accurately with the present choice of $I_{ij}$, 
besides possible systematic errors for the amplitude of order $M/R$
or $v^2/c^2$ where $M$, $R$, and $v$ denote the gravitational mass,
the equatorial circumferential radius, and the radial velocity
of the collapsing star and/or formed neutron stars. 
In stellar core collapses, $v^2/c^2$ is at most $\sim 0.1$, and 
the outcomes are protoneutron stars of $M/R \sim 0.1$.
Thus, it is likely that the wave amplitude is computed within 
$\sim 10\%$ error. The wave phase will be computed 
very accurately as indicated in \cite{SS}. 

\begin{table}[tb]
\begin{center}
\begin{tabular}{|c|c|c|c|c|c|c|c|c|c|} \hline
Model & $\rho_c~({\rm g/cm^3})$ & $M_*(M_{\odot})$
& $M(M_{\odot})$ & $R$~(km) & $T/W$ & $J/M^2$ & $\alpha_c$
& $\Omega_a$ (1/sec) & $\hat A$ \\ \hline
A & $1.00 \times 10^{10}$  & 1.503 & 1.503 & 2267 &
$8.91 \times 10^{-3}$ & 1.235  & 0.994 & 4.11 & $\infty$ \\ \hline
B & $1.00 \times 10^{10}$  & 1.485 & 1.485 & 1576 &
$5.00 \times 10^{-3}$ & 0.839  & 0.994 & 6.49 & 0.32 \\ \hline
C & $1.00 \times 10^{10}$  & 1.488 & 1.488 & 1568 &
$5.44 \times 10^{-3}$ & 0.841  & 0.994 & 8.45 & 1/4 \\ \hline
D & $1.00 \times 10^{10}$  & 1.500 & 1.500 & 1571 &
$1.01 \times 10^{-2}$ & 1.146  & 0.994 & 11.6 & 1/4 \\ \hline
\end{tabular}
\caption{Central density, baryon rest-mass, ADM mass,
equatorial circumferential radius, ratio of the rotational kinetic
energy to the potential energy, non-dimensional angular momentum parameter, 
central value of the lapse function, angular velocity at the
rotational axis, and $\hat A$ of rotating stars chosen as initial
conditions for stellar core collapse simulations. 
}
\end{center}
\vspace{-5mm}
\end{table}

\section{Initial condition and computational setting}

Rotating stellar core
in equilibrium with the $\Gamma=4/3$ polytropic equation of state
[see Eq.(\ref{EOS43})] is given as the initial condition for simulations. 
Following \cite{HD}, the central density is chosen as
$\rho_c = 10^{10}~{\rm g/cm^3}$ irrespective of the velocity profile. 

The velocity profiles of equilibrium rotating stellar cores are given 
according to a popular relation \cite{KEH,Ster} 
\beq
u^t u_{\varphi} = \varpi_d^2( \Omega_a - \Omega ), 
\eeq 
where $\Omega_a$ denotes the angular velocity
along the rotational axis, and $\varpi_d$ is a constant. 
In the Newtonian limit, the rotational profile is written as 
\beq
\Omega = \Omega_a{\varpi_d^2 \over \varpi^2 + \varpi_d^2}. 
\eeq
Thus, $\varpi_d$ indicates the steepness of differential rotation.
In this paper, we pick up the rigidly rotating case in which
$\varpi_d \rightarrow \infty$ (referred to as model A)
and differentially rotating cases 
with $\hat A \equiv \varpi_d/R_e=0.32$ (referred to as model B)
and 1/4 (referred to as models C and D), where $R_e$ is the
coordinate radius at an equatorial surface. In the rigidly rotating case,
we chose the axial ratio of polar radius to equatorial radius as $2/3$.
With this choice, the angular velocity at the equatorial stellar surface
is nearly equal to the Keplerian velocity.
Namely, for the rigidly rotating case, a rapidly rotating initial condition 
with nearly maximum angular velocity is chosen. 
In the differentially rotating case, we chose stars of 
ratio of the rotational kinetic energy $T$ to
the gravitational potential energy $W$ as 
$\sim 0.005$ and $\sim 0.01$ where 
\beqn
T&&={1 \over 2} \int d^3x \rho_* \hat u_{\varphi} \Omega,\\
W&&= \int d^3x \rho_* (1+\varepsilon)-M +T. 
\eeqn
Here, $W$ is defined to be positive. 
In Table I, several quantities for the models adopted
in the present numerical computation are summarized. 

For the differentially rotating case with a small value of $\hat A(< 1)$,
it is possible to make equilibrium states with $T/W \gg 0.01$. 
With such an initial condition, the collapsing stellar core often 
forms a differentially rotating star 
of a highly nonspherical shape and of a high value of $T/W$ \cite{Newton45}. 
It is also known that rapidly rotating neutron stars of a high degree of 
differential rotation is dynamically unstable against nonaxisymmetric 
deformation (e.g., \cite{SKS} and references therein). 
To simulate the collapse with a high initial value of 
$T/W$, a nonaxisymmetric simulation will be necessary. 
Since our attention here is restricted to the axisymmetric case,
we do not choose such initial conditions. 

Simulations are performed for four initial conditions listed in
Table I. Models A and B are almost the same initial conditions as 
models A1B3 and A3B2 in \cite{HD}.
(Note that the value of $A$ for model B is $\approx 5 \times 10^2$ km 
which is approximately equal to that for model A3B2 in \cite{HD}.) 
Careful comparison of present numerical results with those in \cite{HD} 
is carried out using these two models. A variety of values of 
$\Gamma_1$, $\Gamma_2$, and $\Gamma_{\rm th}$ are adopted to
investigate the dependence of numerical results on the equations 
of state: $\Gamma_1$ is chosen 
as 1.28, 1.30, 1.31, and 1.32, $\Gamma_2$ as 2 and 2.5, and 
$\Gamma_{\rm th}$ as 1.35, 1.5, and 5/3. 
The selected sets are listed in Table II. 

\begin{table}[tb]
\begin{center}
\begin{tabular}{|c|c|c|c|} \hline
Model & $\Gamma_1$ & $\Gamma_2$ & $\Gamma_{\rm th}$ 
\\ \hline
A1 & 1.32 & 2.5  & 1.5 \\ \hline
A2 & 1.31 & 2.5  & 1.5 \\ \hline
A3 & 1.28 & 2.5  & 1.5 \\ \hline
A4 & 1.32 & 2.5  & 1.35 \\ \hline
A5 & 1.32 & 2.5  & 5/3  \\ \hline
A6 & 1.32 & 2.0  & 1.5  \\ \hline
B1 & 1.32 & 2.5  & 1.5 \\ \hline
B2 & 1.30 & 2.5  & 1.5 \\ \hline
C1 & 1.32 & 2.5  & 1.5 \\ \hline
C2 & 1.30 & 2.5  & 1.5 \\ \hline
C3 & 1.32 & 2.0  & 1.5 \\ \hline
C4 & 1.32 & 2.5  & 1.35 \\ \hline
D  & 1.32 & 2.5  & 1.5 \\ \hline
\end{tabular}
\caption{Selected sets of $\Gamma_1$, $\Gamma_2$, and $\Gamma_{\rm th}$. 
}
\end{center}
\vspace{-5mm}
\end{table}

\begin{table}[tb]
\begin{center}
\begin{tabular}{|c|c|c|c|c|} \hline
Model &  $\Delta x$ (initial) & $L$ (initial) & $\Delta x$ (final)
& $L$ (final) \\ \hline
A & 3.775 & 2265 & 0.4719 & 1180 \\ \hline
A(low resolution) & 6.292 & 2643 & 0.7865 & 1337 \\ \hline
B & 2.601 & 1613 & 0.3251 & 813 \\ \hline 
B(low resolution) & 3.902 & 1639 & 0.4877 & 829 \\ \hline 
C & 2.613 & 1568 & 0.3267 & 817 \\ \hline
D & 2.617 & 1570 & 0.3271 & 818 \\ \hline
\end{tabular}
\caption{The initial and final grid spacings and location of the outer
boundaries along the $x$ and $z$ axes for models A--D. The unit is
kilometer.
}
\end{center}
\vspace{-5mm}
\end{table}

During the collapse, the central density increases from
$10^{10}~{\rm g/cm^3}$ to $\sim 5 \times 10^{14}~{\rm g/cm^3}$.
This implies that the characteristic length scale of the system varies by
a factor of $\sim 100$. One of the computational issues in a stellar core
collapse simulation is to guarantee numerical accuracy against the 
significant change of the characteristic length scale. 
In the early phase of the collapse (infall phase; see Sec. IV A) in which 
it proceeds in a nearly homologous manner, we may follow the collapse 
with a relatively small number of grid points by moving the outer
boundary inward while decreasing the grid spacing, 
without increasing the grid number by a large factor. 
As the collapse proceeds, the central region shrinks more rapidly than 
the outer region does and, hence, a better grid resolution is 
necessary to accurately follow such a rapid collapse in the central region. 
On the other hand, the location of the outer boundaries cannot be changed 
by a large factor to avoid discarding the matter in the outer 
envelopes. 

To compute such a collapse accurately while saving the CPU time efficiently,
a regridding technique as described in \cite{SS02} is adopted. 
The regridding is carried out whenever the characteristic radius
of the collapsing star decreases by a factor of a few. At each regridding, 
the grid spacing is decreased by a factor of 2. 
All the quantities in the new grid are calculated
using the cubic interpolation. 
To avoid discarding the matter in the outer region, we also increase the 
grid number at the regridding, keeping a rule that 
the discarded baryon rest-mass has to be less than 3\% of the total. 

Specifically, $N$ and $L$ in the present work
are chosen in the following manner. First, we define 
a relativistic gravitational potential
$\Phi_c \equiv 1 -\alpha_c~ (\Phi_c>0)$, which  
is $\approx 0.006$ at $t=0$ for all the models chosen in this work. 
Since $\Phi_c$ is approximately proportional to $M/R$, 
$\Phi_c^{-1}$ can be used as a measure of the characteristic
length scale for the regridding. From $t=0$ to the time at which
$\Phi_c = 0.025$, we set $N=620$. Note that 
the equatorial radius is initially covered by 600 grid points.
At $\Phi_c = 0.025$, the characteristic stellar radius
becomes approximately one fourth of the initial value. 
Then, the first regridding is performed; the grid 
spacing is changed to the half of the previous one
and the grid number is increased to $N=1020$. 
Subsequently, the value of $N$ is chosen in the following manner; 
for $0.025 \leq \Phi_c \leq 0.05$, we set $N=1020$; 
for $0.05 \leq \Phi_c \leq 0.1$, we set $N=1700$; and 
for $0.1 \leq \Phi_c $, we set $N=2500$ and keep 
this number until the termination of the simulations 
since the maximum value of $\Phi_c$ is at most 0.25.
In this treatment, the total discarded fraction of the baryon rest-mass
which is located outside new regridded domains is $\alt 3\%$. 

To check the convergence of numerical results,
we also perform a few simulations 
using a low grid resolution (cf. Table III).
In this case, the value of $N$ is changed as follows:
for $\Phi_c \leq 0.025$, $N=420$; 
for $0.025 \leq \Phi_c \leq 0.05$, $N=820$; 
for $0.05 \leq \Phi_c \leq 0.1$, $N=1300$; and 
for $0.1 \leq \Phi_c $, $N=1700$. 

Simulations for each model with the higher grid resolution 
are performed for 60000--80000 time steps. 
The required CPU time for one model is about 40--70 hours using 8
processors of FACOM VPP 5000 at the data processing center
of National Astronomical Observatory of Japan. 

\section{Numerical Results}

\subsection{Dynamics of the collapse}

\subsubsection{General feature}

\begin{figure}[htb]
%\vspace*{-4mm}
\begin{center}
\epsfxsize=3.1in
\leavevmode
(a)\epsffile{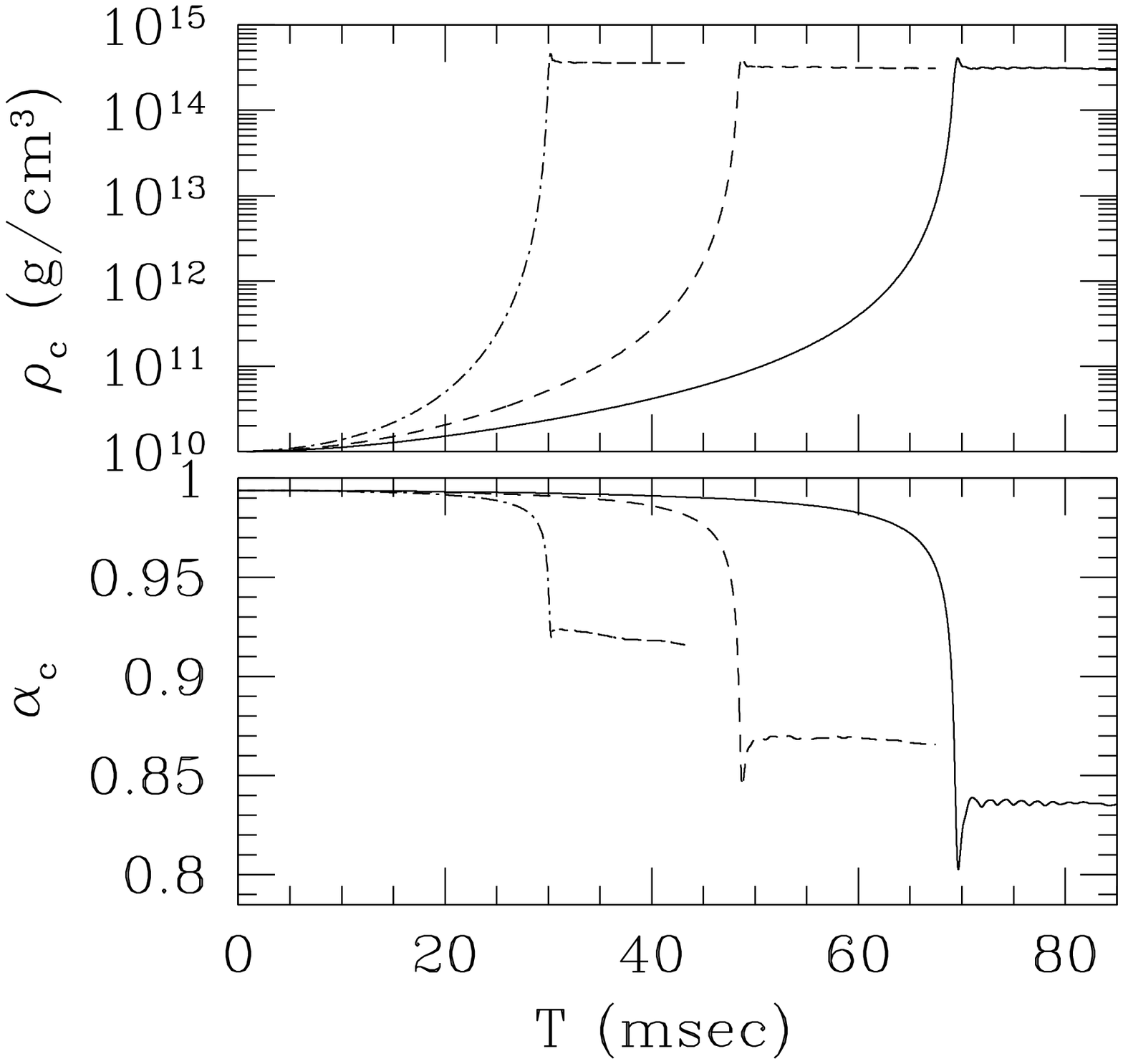}
\epsfxsize=3.1in
\leavevmode
(b)\epsffile{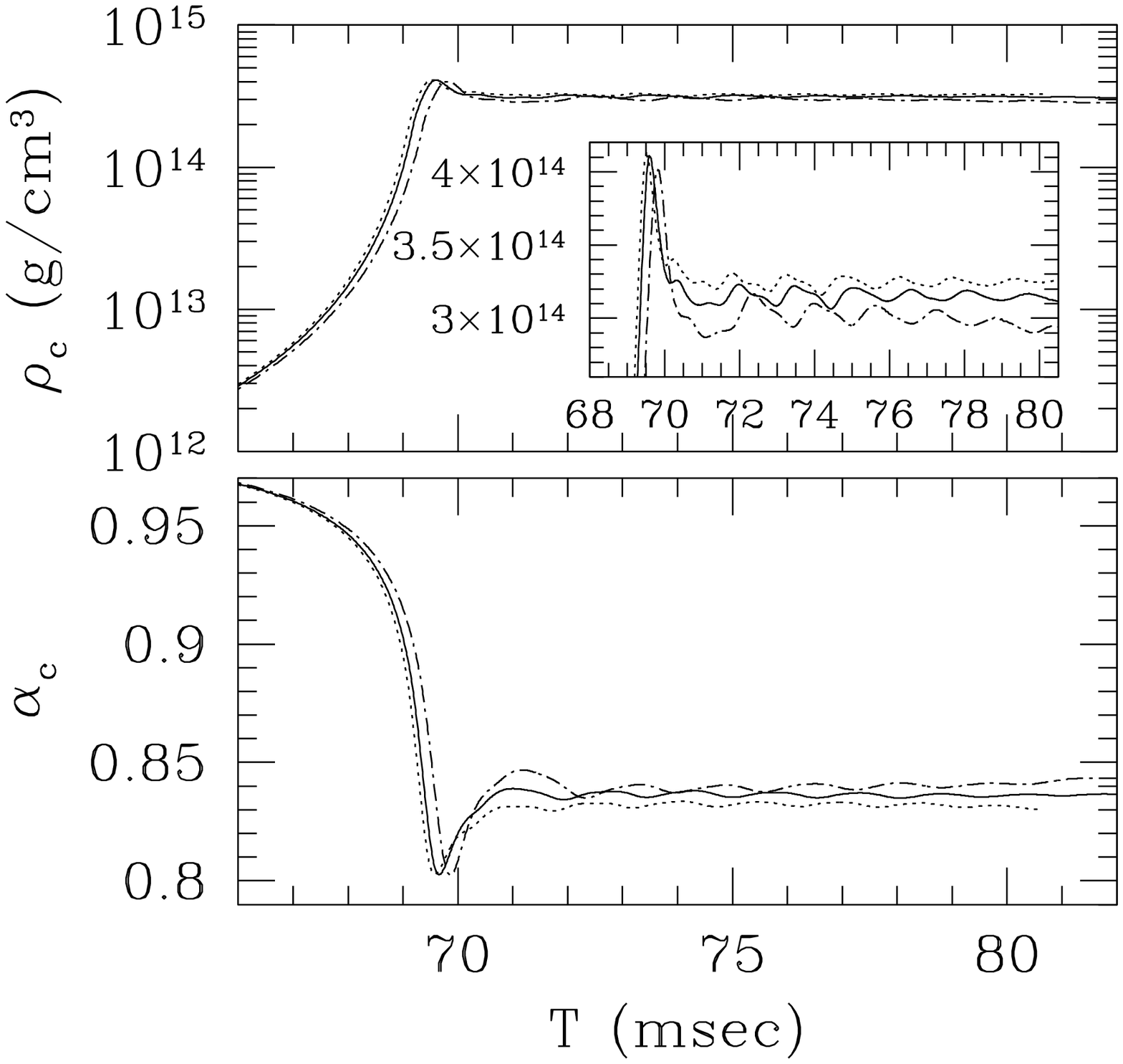} \\
\epsfxsize=3.1in
\leavevmode
(c)\epsffile{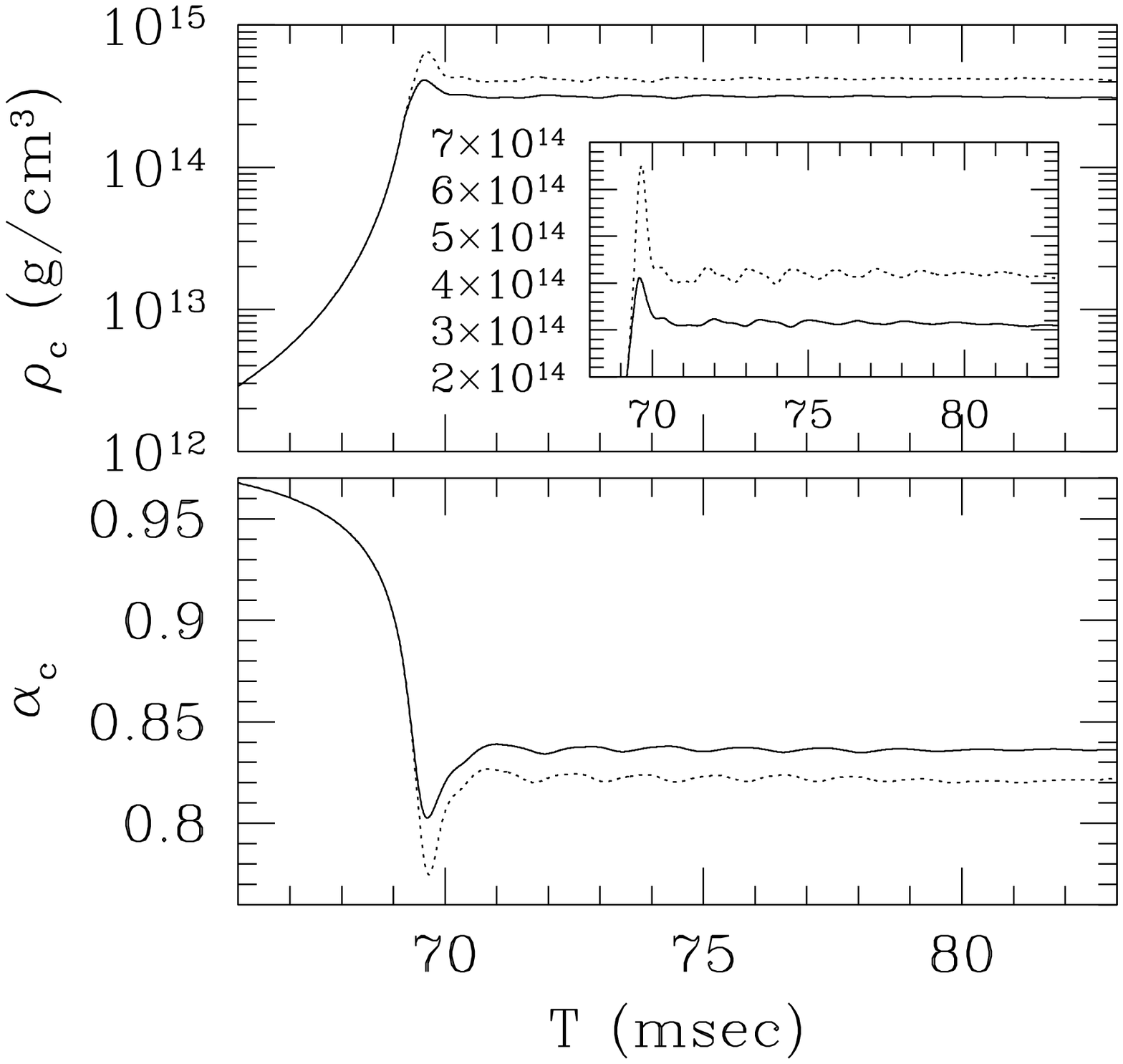}
\epsfxsize=3.1in
\leavevmode
(d)\epsffile{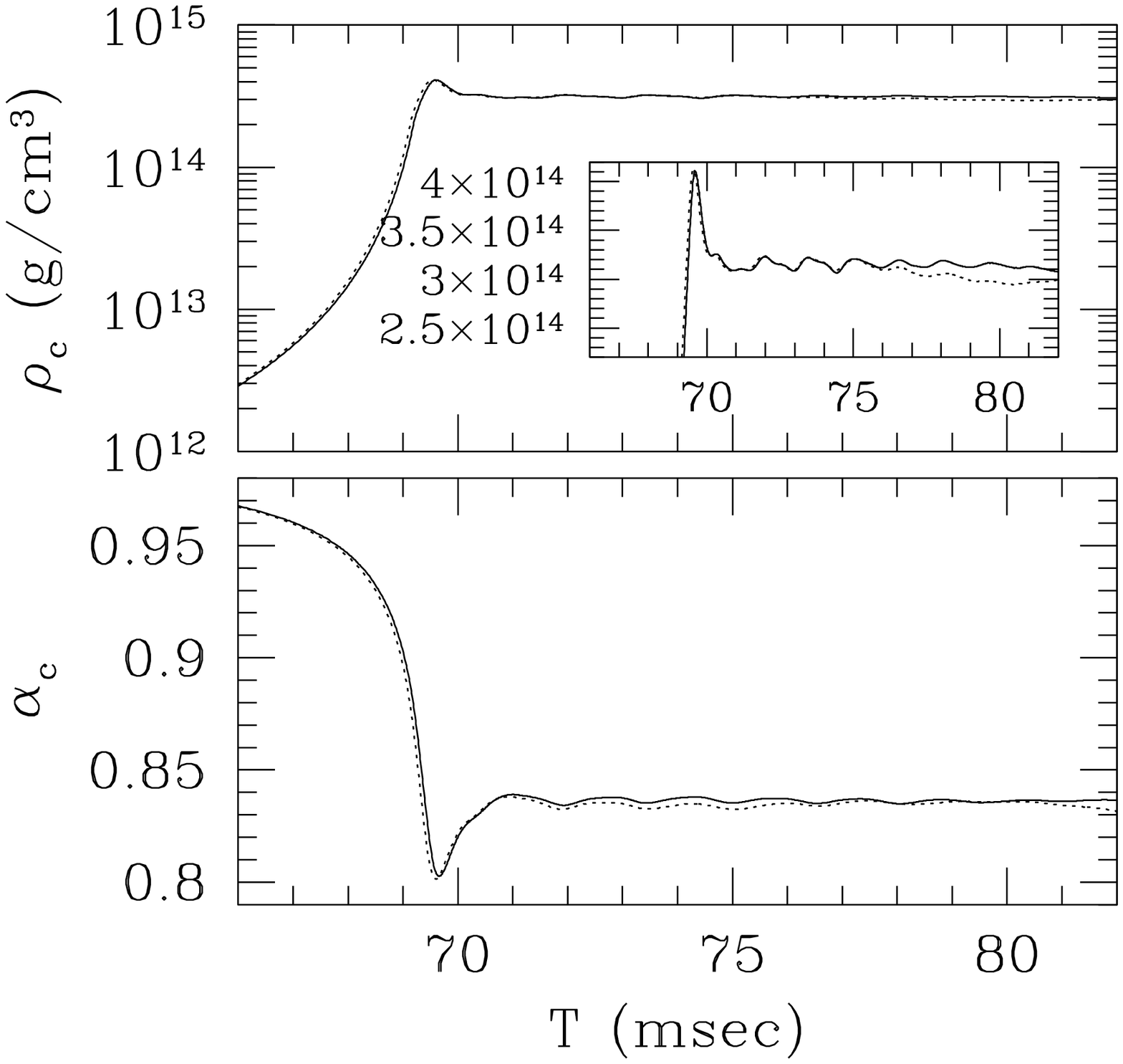}
\caption{Evolution of the central density and the central value of 
the lapse function for model A. 
(a) Models A1 (solid curve), A2 (dashed curve), and A3 (dotted-dashed curve); 
(b) Models A1 (solid curve), A4 (dotted curve), and A5 (dotted-dashed curve);
(c) Models A1 (solid curve) and A6 (dotted curve); 
(d) Model A1 with high (solid curve) and low grid resolutions (dotted curve). 
\label{FIG1}
}
\end{center}
\end{figure}

\begin{figure}[htb]
%\vspace*{-4mm}
\begin{center}
\epsfxsize=3.1in
\leavevmode
(a)\epsffile{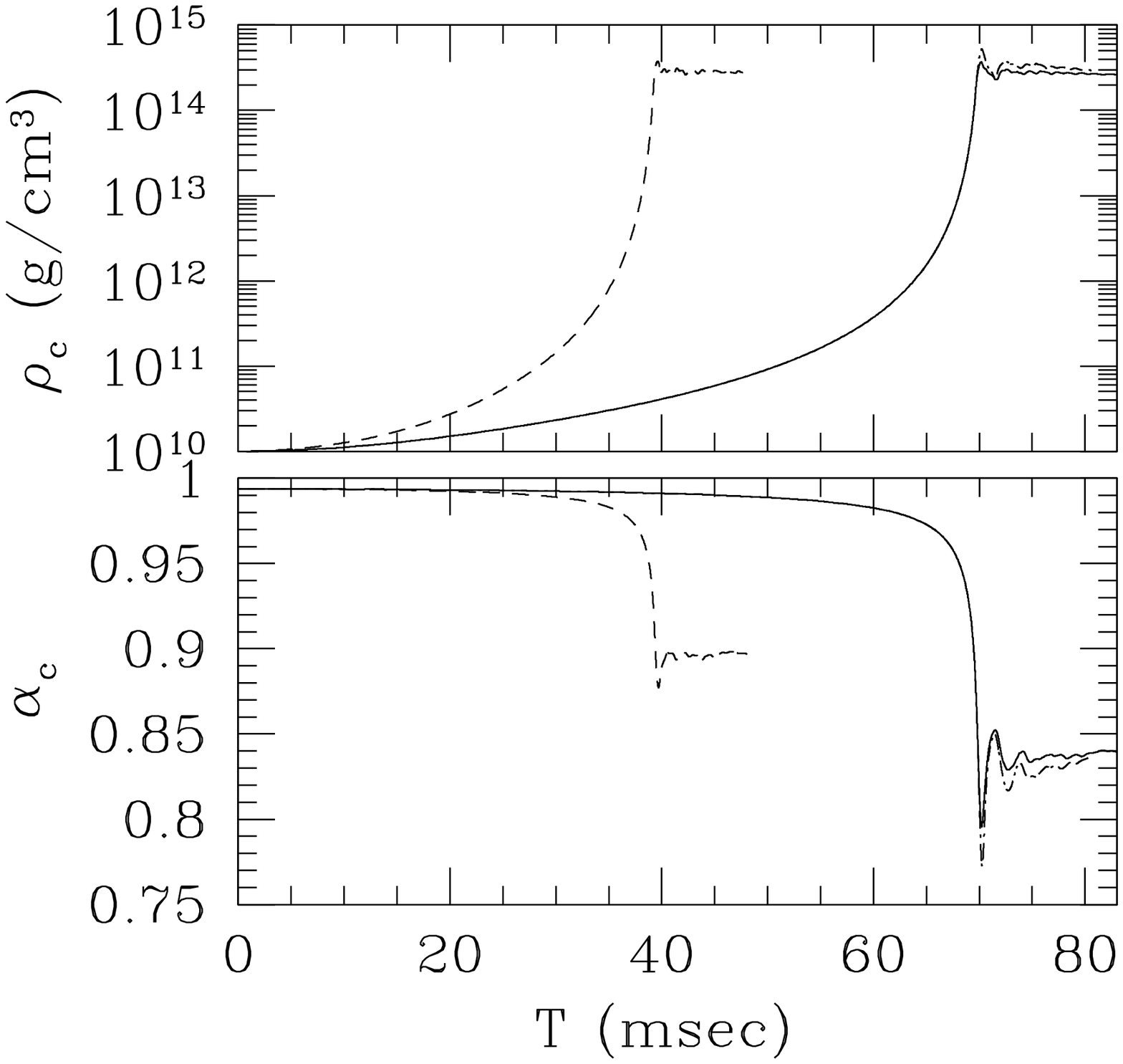}
\epsfxsize=3.1in
\leavevmode
(b)\epsffile{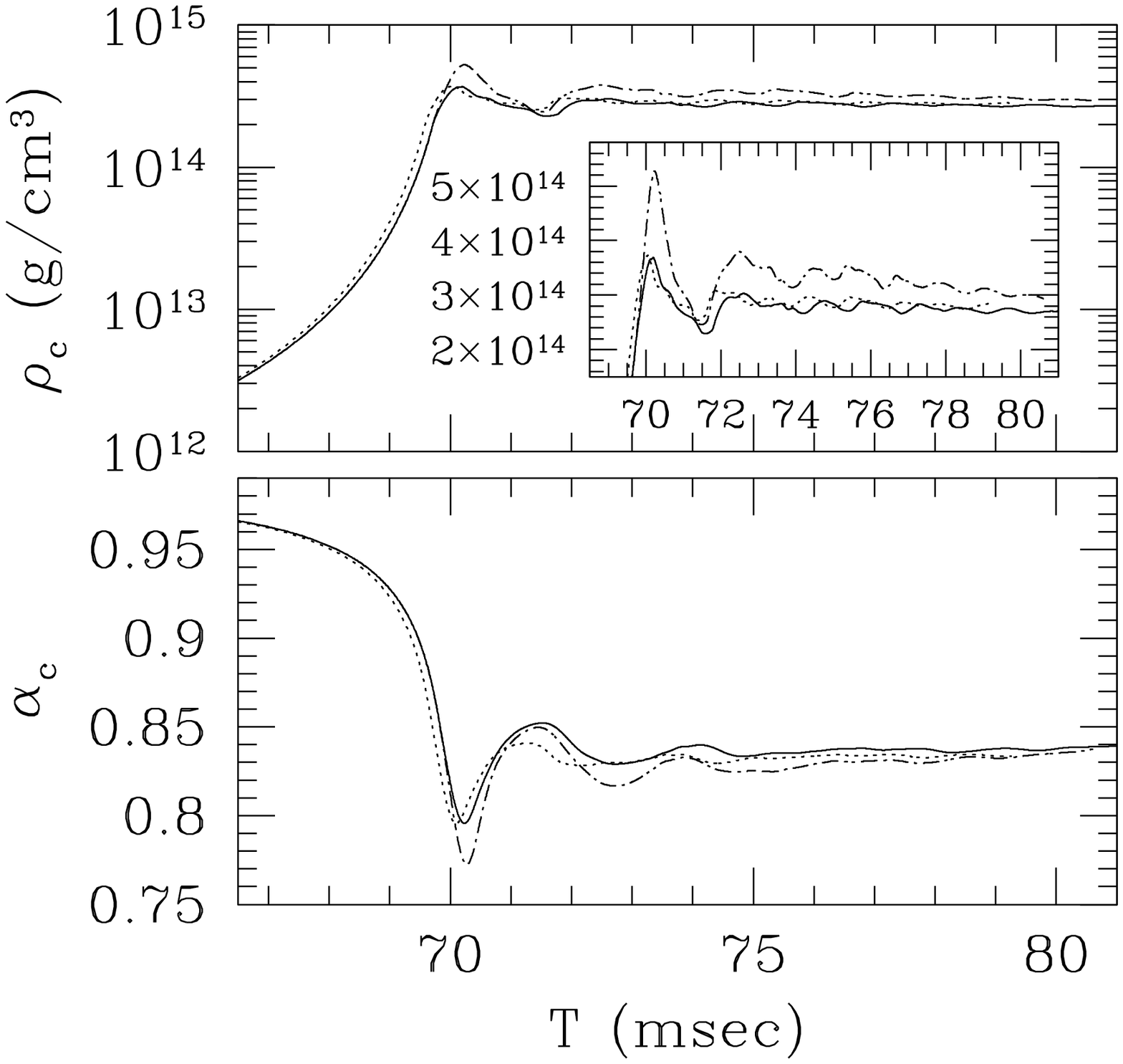}
%\vspace*{-4mm}
\caption{
(a) The same as Fig. 1 but for models C1 (solid curves),
C2 (dashed curves), and C3 (dotted-dashed curves).
(b) The same as Fig. 1 but for models C1 (solid curves), 
C3 (dotted-dashed curves), and C4 (dotted curves). 
\label{FIG2}
}
\end{center}
\end{figure}

\begin{figure}[htb]
%\vspace*{-4mm}
\begin{center}
\epsfxsize=2.95in
\leavevmode
(a)\epsffile{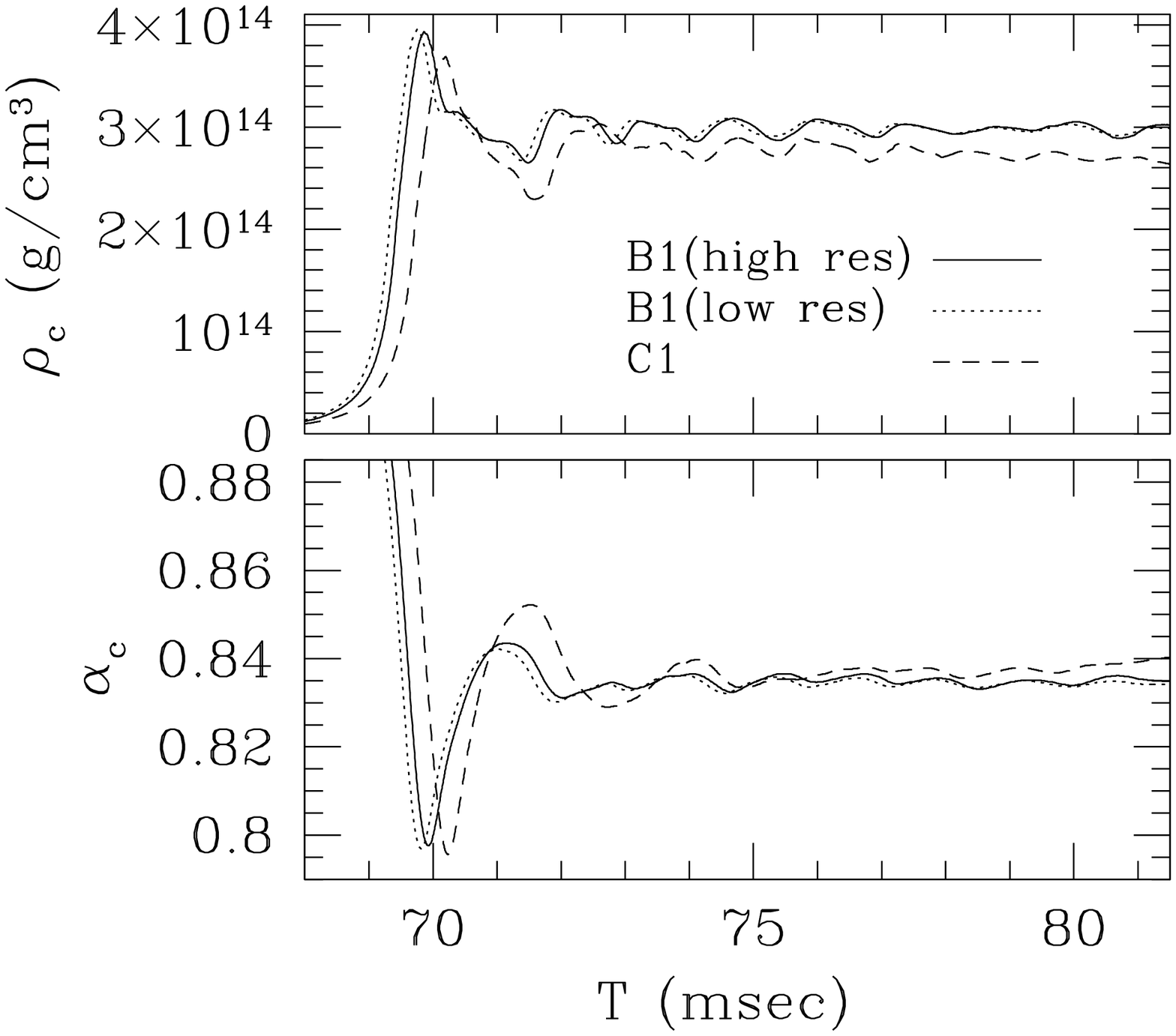}
\epsfxsize=2.95in
\leavevmode
(b)\epsffile{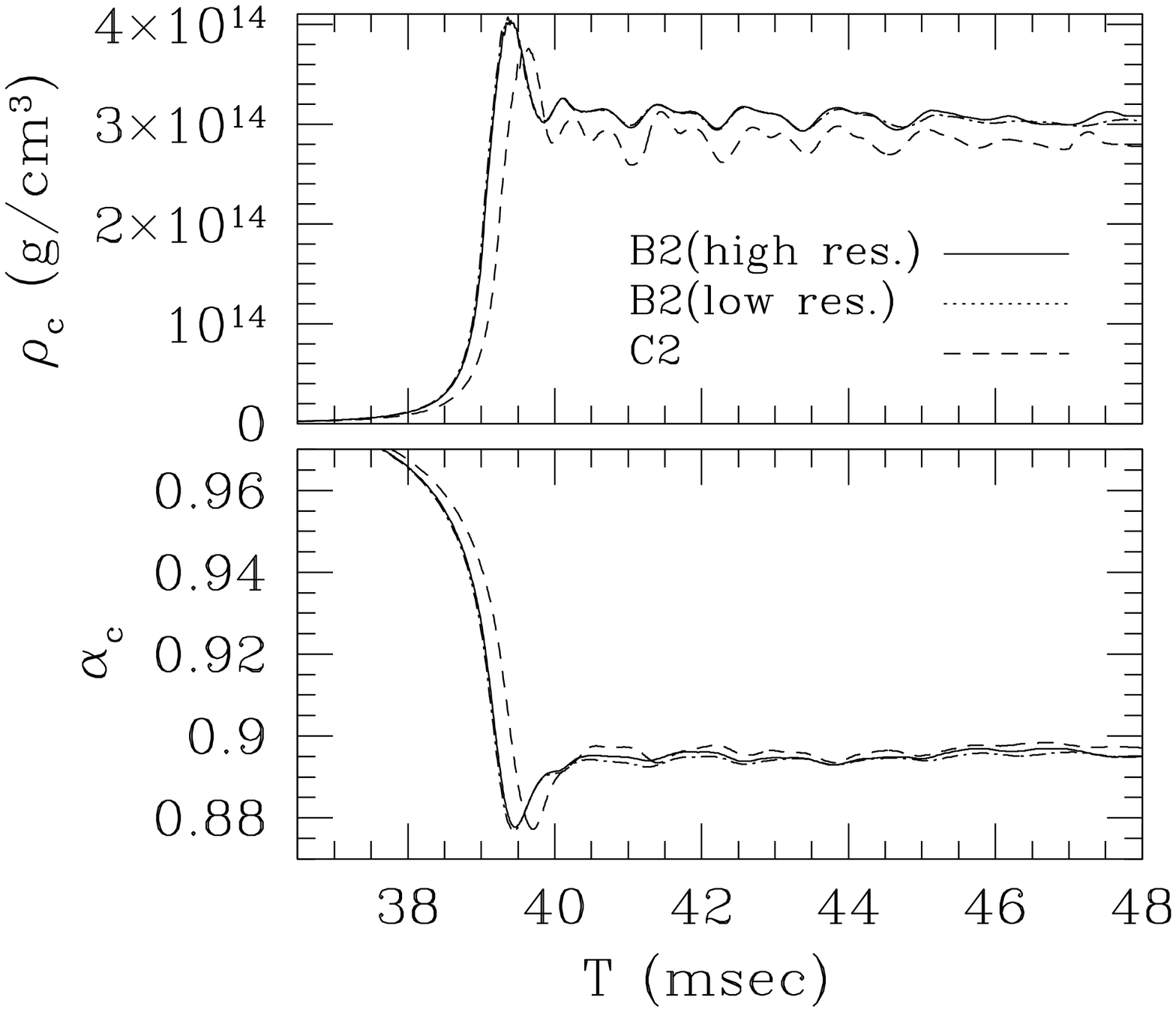}
%\vspace*{-4mm}
\caption{
(a) The same as Fig. 1 but for model B1 with high 
(solid curve) and low grid resolutions (dotted curve). For comparison, 
the results for model C1 (dashed curve) are also shown. 
(b) The same as (a) but for model B2 with high 
(solid curve) and low grid resolutions (dotted curve) 
and for model C2 (dashed curve).  
\label{FIG3}
}
\end{center}
\end{figure}

\begin{figure}[htb]
%\vspace*{-4mm}
\begin{center}
\epsfxsize=2.95in
\leavevmode
\epsffile{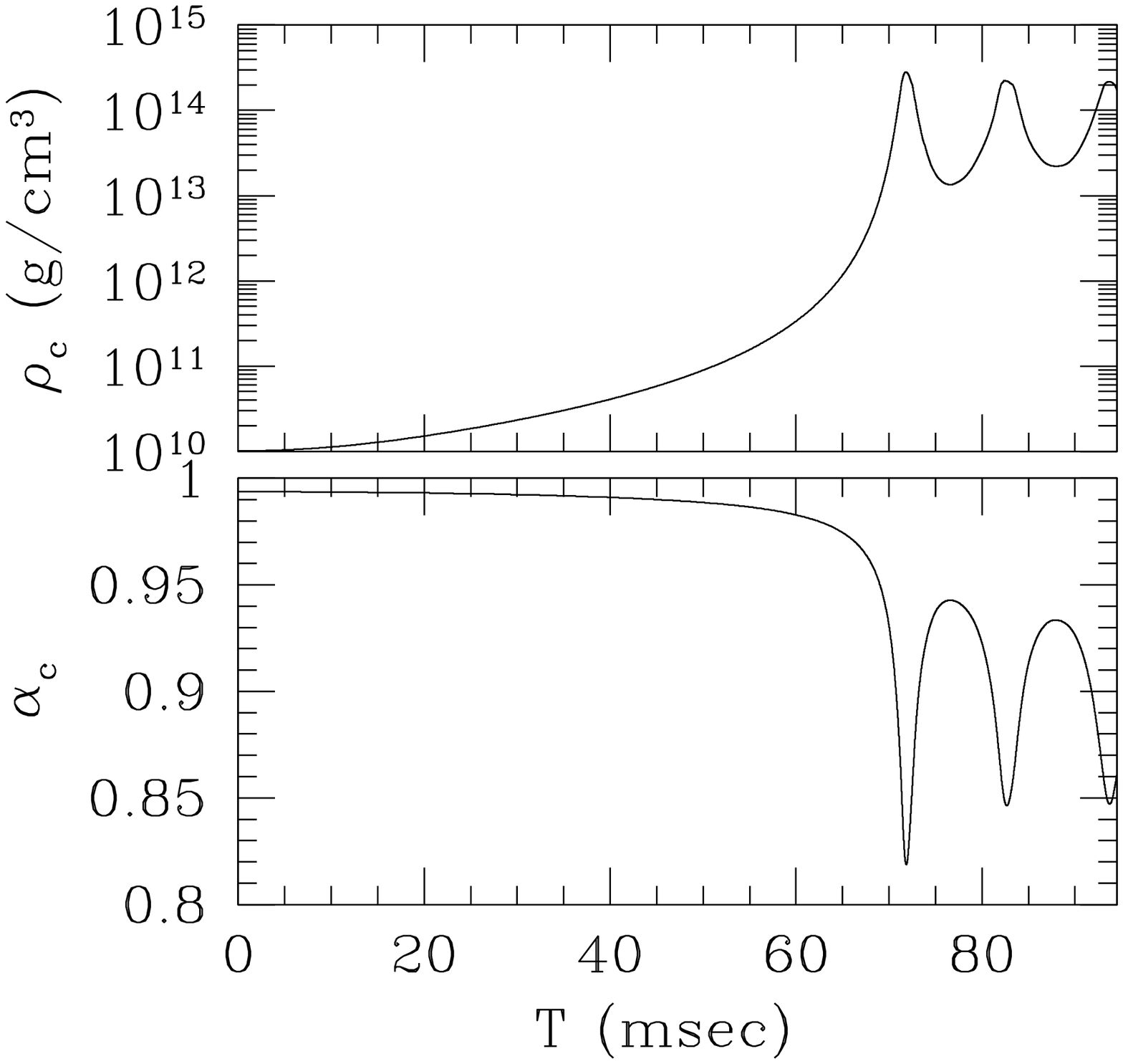}
\vspace*{-4mm}
\caption{Evolution of the central density and the central value
of the lapse function for model D. 
\label{FIG4}
}
\end{center}
\end{figure}

Figures 1--4 show the evolution of the central density (hereafter $\rho_c$)
and the central value of the lapse function (hereafter $\alpha_c$) 
for models A -- D. Figures 5 and 6 are snapshots of 
the density contour curves and the velocity vectors of $(v^x, v^z)$
on the $y=0$ plane for models A1 and C1
at selected time slices around which shocks are formed. 

As described in \cite{HD}, rotating stellar core collapses
can be divided into three phases.
The first one is the infall phase in which the 
core collapse proceeds from the onset of the
gravitational instability triggered by the sudden softening of
the equation of state due to the reduction of the adiabatic
index. During this phase, the central density (the central
value of the lapse function) monotonically increases (decreases)
until it reaches the nuclear density or the centrifugal force
becomes strong enough to halt the collapse. 
The inner part of the core, which collapses nearly homologously,
constitutes the inner core. 
The duration of the infall phase in the present work
is between about 30 and 70 msec depending mainly
on the value of $\Gamma_1$ 
(for the smaller value of $\Gamma_1$, the duration is shorter)
as shown in \cite{HD}. We note that the dynamical time at
$t=0$ defined by $\rho_c^{-1/2}$ is $\approx 38.7$ msec. Thus,
the duration may be written as 0.8--1.8 $\rho_c^{-1/2}$. 

The second one is the bounce phase which sets in when the densities
around the central part exceed the nuclear density $\rho_{\rm nuc}$, or
when centrifugal forces, which become stronger as
the collapse proceeds due to angular momentum conservation,  
begin to dominate over gravitational attraction force.  
At this phase, the inner core decelerates infall 
in about a few msec ($\sim 10\rho_{\rm nuc}^{-1/2}$).
Because of its large inertia and large kinetic
energy due to the infall, the inner core does not settle down
to a stationary state immediately but overshoots and bounces
back, forming shocks at the outer edge of the inner core. 

The third one is the ring-down phase or the re-expansion phase.
If the centrifugal force is sufficiently small at the time that 
the density of the inner core exceeds the nuclear density, 
the bounce occurs when the central density reaches
$\sim 2$--$3\rho_{\rm nuc}$ due to a sudden stiffening of the 
equation of state. In this case, the inner core quasiradially oscillates
for about 10 msec and then settles down to a quasistationary state.
In the outer region, on the other hand, shock waves propagate outward 
sweeping materials which infall from outer envelopes. 

If the angular momentum in the inner region is sufficiently large,
the collapse is halted by the centrifugal force, not
by the sudden stiffening of the equation of state. In this case, the 
stellar core does not settle down to a quasistationary state.
Instead, it rebounds due to the centrifugal force and expands to 
be of a subnuclear density. After the maximum expansion is reached, 
the core starts collapsing again.
It repeats the bounce, the expansion, and the collapse for many times. 
During each bounce, shocks are formed at the outer region of
the core and the 
oscillation amplitude is damped gradually due to the shock dissipation. 

%%Eventually, the density of the core will settle down to a 
%%subnuclear density. Since the equation of state for the rotating star 
%%of subnuclear matter is very soft, it will be stabilized by the large
%%centrifugal force. 

For models A--C, the centrifugal force 
is not strong enough to halt the collapse and, hence, 
a protoneutron star of the central density larger than the
nuclear density is formed irrespective of the values
of $\Gamma_1$, $\Gamma_2$, and $\Gamma_{\rm th}$ (see Figs. 1--3).
On the other hand, for model D, the angular momentum is large enough
to halt the collapse and to prevent the inner core being compact.
As a result, the outcome is an oscillating star of a 
subnuclear density (see Fig. 4). Since the amplitude of the 
oscillation decreases gradually, it will settle to a 
rotating star of subnuclear density eventually. 
The adiabatic constant of this star is $\approx \Gamma_1$ that is 
smaller than 4/3 which is the well-known critical value
against gravitational collapse for spherical stars, and $1.329$ which is 
an approximate critical value for rigidly rotating stars \cite{shiba03}. 
This indicates that the centrifugal force by a rapid and
differential rotation plays an essential role for the stabilization
against gravitational collapse.
According to \cite{Tas}, the criterion of the stability
for slowly rotating stars is given by 
\beq
Q_c \equiv
3\Gamma_1 - 4 - 2 {T \over W}(3\Gamma_1 - 5) -k {M \over R} > 0,
\eeq
where $k$ is a constant which is $\approx 6.75$ for $n=3$ and
$T/W=0$ \cite{Ch64}.
For the Newtonian polytropes with $n \approx 3$, the stellar radius is
given by
\beq
R \approx 2.35 \biggl({M \over \rho_c}\biggr)^{1/3}
\approx 73~{\rm km} \biggl({M \over 1.5M_{\odot}}\biggr)^{1/3}
\biggl({\rho_c \over 10^{14}~{\rm g/cm^3}}\biggr)^{-1/3}. 
\eeq
Thus, $M/R$ will be $\sim 0.03$ for $\rho_c \sim 10^{14}~{\rm g/cm^3}$. 
The value of $T/W$ for dynamical stars is not exactly 
defined in general relativity, but assuming that 
it approximately increases as $1/R \propto 1- \alpha_c$
for a fixed value of $M$, 
we can infer that the value of $T/W$ would be 
$\sim 0.15$--0.2 for $\alpha_c \sim 0.9$ and $k=6.75$.
Therefore, $Q_c$ would be $\sim 0.1$--0.2, 
and, hence, the rotating star would satisfy the stability condition
against gravitational collapse. 
On the other hand, the expected value of $T/W$ 
is so large that the formed differentially rotating 
star may be unstable against a nonaxisymmetric deformation \cite{SKS}. 
This suggests that to clarify the fate of this star, it would 
be necessary to perform a nonaxisymmetric simulation \cite{SBS}. 
However, such a simulation is beyond scope of this paper and, hence,
particular attentions are paid only to models A--C in this paper. 

As Figs. 1(a) and 2(a) indicate, the evolution of the 
central density and the central value of the lapse function
depends strongly on the value of $\Gamma_1$. 
%%even when values of $\Gamma_2(=2.5)$ and
%%$\Gamma_{\rm th}(=1.5)$ are fixed.
For the smaller value of $\Gamma_1$, 
the depleted pressure at $t=0$ is larger. As a result, 
the collapse is accelerated more and the elapsed time in the infall phase 
is shorter. Also, since the depleted fraction of the pressure is larger
in the central region than in the outer region, the collapse
in the central region proceeds more rapidly. This results in 
a less coherent collapse for the smaller value of $\Gamma_1$. 
This effect makes the mass of a protoneutron star at its formation 
smaller and is reflected in the value of $\alpha_c$ in the ring-down
phase which depends on the compactness of the protoneutron star. 
On the other hand, the final value of $\rho_c$ depends only weakly
on the value of $\Gamma_1$. This indicates that for the smaller
value of $\Gamma_1$, the formed protoneutron star has a more
centrally concentrated structure. 

In Figs. 1(b) and 2(b), the evolution of the 
central density and the central value of the lapse function 
for different values of $\Gamma_{\rm th}$ 
with fixed values of $\Gamma_1(=1.32)$ and $\Gamma_2(=2.5)$ is compared. 
Recall that the value of $\Gamma_{\rm th}$ determines the strength of shocks 
at the bounce and at their subsequent propagation. 
Thus, the results here show that a moderate 
change of the value of $\Gamma_{\rm th}$ from 1.35 to 5/3 weakly 
modifies the evolution of the formed protoneutron stars. 
For the smaller value of $\Gamma_{\rm th}$, 
the final value of the central density (central lapse) is larger (smaller),
This is simply because the amount of matter that accretes to 
the protoneutron star increases and, hence, the compactness
increases with the decrease of the value of $\Gamma_{\rm th}$. 
%%%due to a weaker effect of the shocks. 
For the larger value of $\Gamma_{\rm th}$, 
the oscillation amplitude of $\rho_c$ is larger. 
This is due to the fact that the stronger shocks
result in the larger amplitude of the oscillation of the core. 

In Figs. 1(c) and 2(b), the evolution of the central density
and the central value of the lapse function is compared 
for different values of $\Gamma_{2}$ with fixed values of 
$\Gamma_1(=1.32)$ and $\Gamma_{\rm th}(=1.5)$. 
[Compare the solid and dotted-dashed curves in Fig. 2(b).] 
Since the equation of state for a protoneutron star is stiffer 
for the larger value of $\Gamma_2$, the maximum density at 
the bounce, the final relaxed value of $\rho_c$, and 
the compactness of the quasistationary neutron star are smaller.
Since the infall proceeds deeply inside the core, 
the amplitude of the oscillation for the central density
in the ring-down phase is larger for the smaller value of $\Gamma_2$. 

Figure 3 shows the evolution of the central density and
the central value of the lapse function
(a) for models B1 and C1 and (b) for models B2 and C2.
The values of $\Gamma_1$, $\Gamma_2$, and $\Gamma_{\rm th}$ 
are identical between models B1 and C1 and between models B2 and C2. 
Furthermore, the values of $T/W$ for the initial condition are
approximately equal. Therefore, the difference of the numerical results
comes from the angular velocity profile of the initial conditions.
Figure 3 indicates that the degree of differential rotation at $t=0$ 
is reflected significantly in the oscillation and 
evolution of the formed protoneutron stars. 
The quantitative differences are summarized as follows: 
(i) the time at the bounce $t_b$ for models C1 and C2 is slightly
larger than that for models B1 and B2, respectively; 
(ii) the maximum value of the central density for models C1 and C2
is slightly smaller than that for models B1 and B2, respectively; 
(iii) the amplitude of the oscillation 
of the central density and central value of the lapse function
in the ring-down phase is larger for models C1 and C2. The results 
(i) and (ii) are simply due to the fact that the centrifugal 
force around the central region for models C1 and C2 is slightly larger 
and plays a stronger role for halting the collapse. 
The result (iii) indicates that the small increase of the angular velocity
around the central region in the initial condition
can significantly modify the evolution of the central density. 
All the results (i)--(iii) also show that the oscillation of the
central density of the formed protoneutron stars depends strongly
on the initial angular velocity profile. 

Effects of differential rotation of the initial condition 
are also reflected significantly in the shape of the 
formed protoneutron stars. In the collapse of 
a rigidly rotating progenitor, the formed protoneutron star has 
a slightly nonspherical shape (see Fig. 5). On the other hand, 
in the collapse of a differentially rotating progenitor,
a protoneutron star of a flattened and nonspherical shape is
the outcome (see Fig. 6). This difference results from the fact that
the inner region is more rapidly rotating in the case of the
differentially rotating progenitor. 
It is worthy to note that the value of $T/W$ for model A is about 1.6 times 
as large as that for model C. However, the angular velocity 
at the rotational axis for model A is about half of that for model C.
Thus, $T/W$ alone is not a good indicator for 
measuring the significance of the centrifugal force 
in rotating stellar core collapses (nor is the 
nondimensional angular momentum parameter $J/M^2$). 
Obviously, the local distribution of the angular momentum
plays a more important role for determining the shapes of
the formed protoneutron star and shocks. 

Convergence of the numerical results is achieved well in the
present computation. In Figs. 1(d), 3(a), and 3(b),
we show the numerical results with a low grid
resolution for models A1, B1, and B2 (dotted curves).
It is found that the evolution of the
central density and the central lapse in the low-resolution
simulation agrees with that in the high-resolution one
within a small error 
(except for the very late time for which the numerical error
seems to be accumulated for the low-resolution simulation). 
This indicates that the grid resolutions adopted in the present
numerical simulation are fine enough to yield a convergent
numerical result. 

\begin{figure}[htb]
%\vspace*{-4mm}
\begin{center}
\epsfxsize=2.2in
\leavevmode
\epsffile{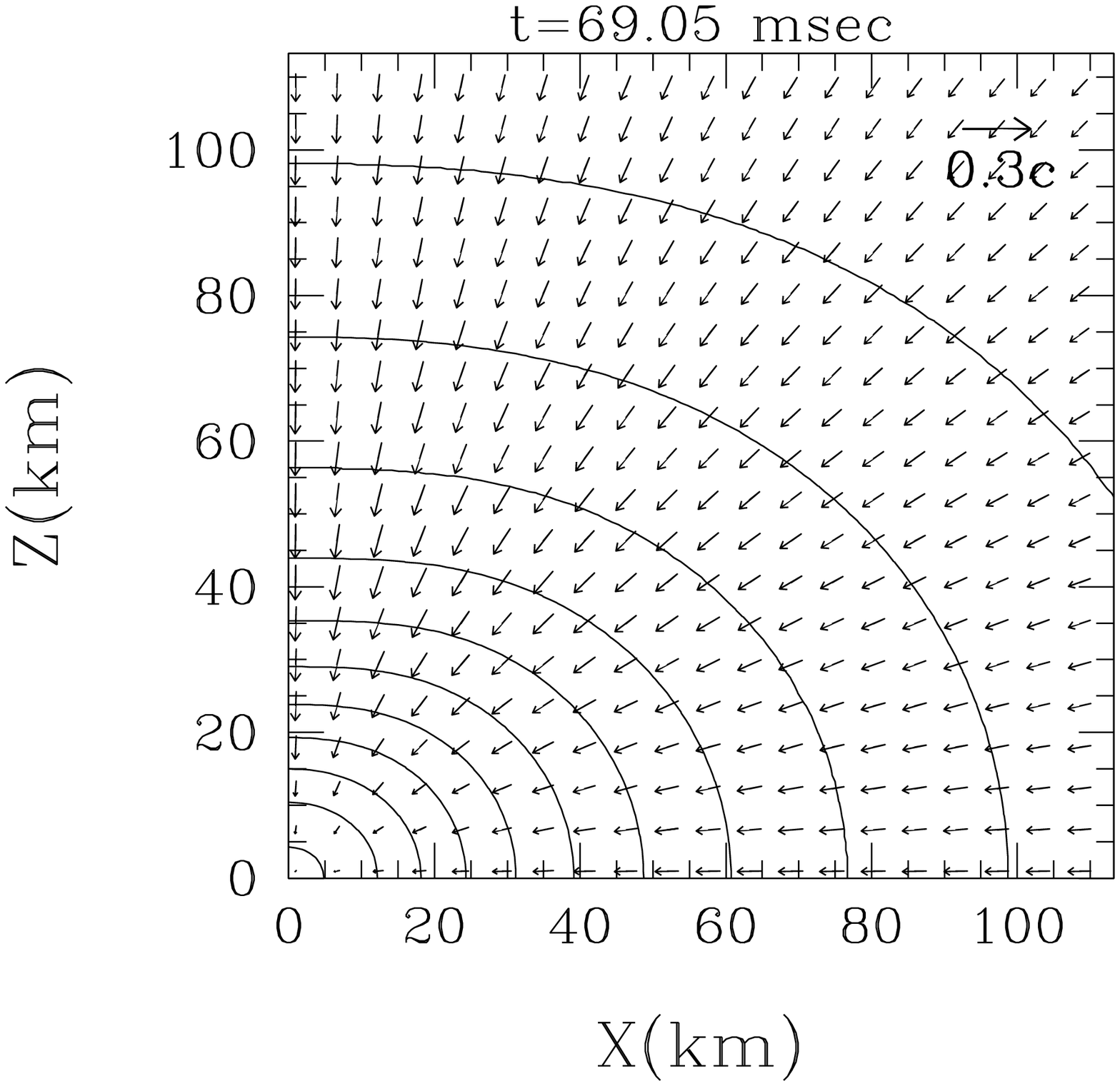}
\epsfxsize=2.2in
\leavevmode
\epsffile{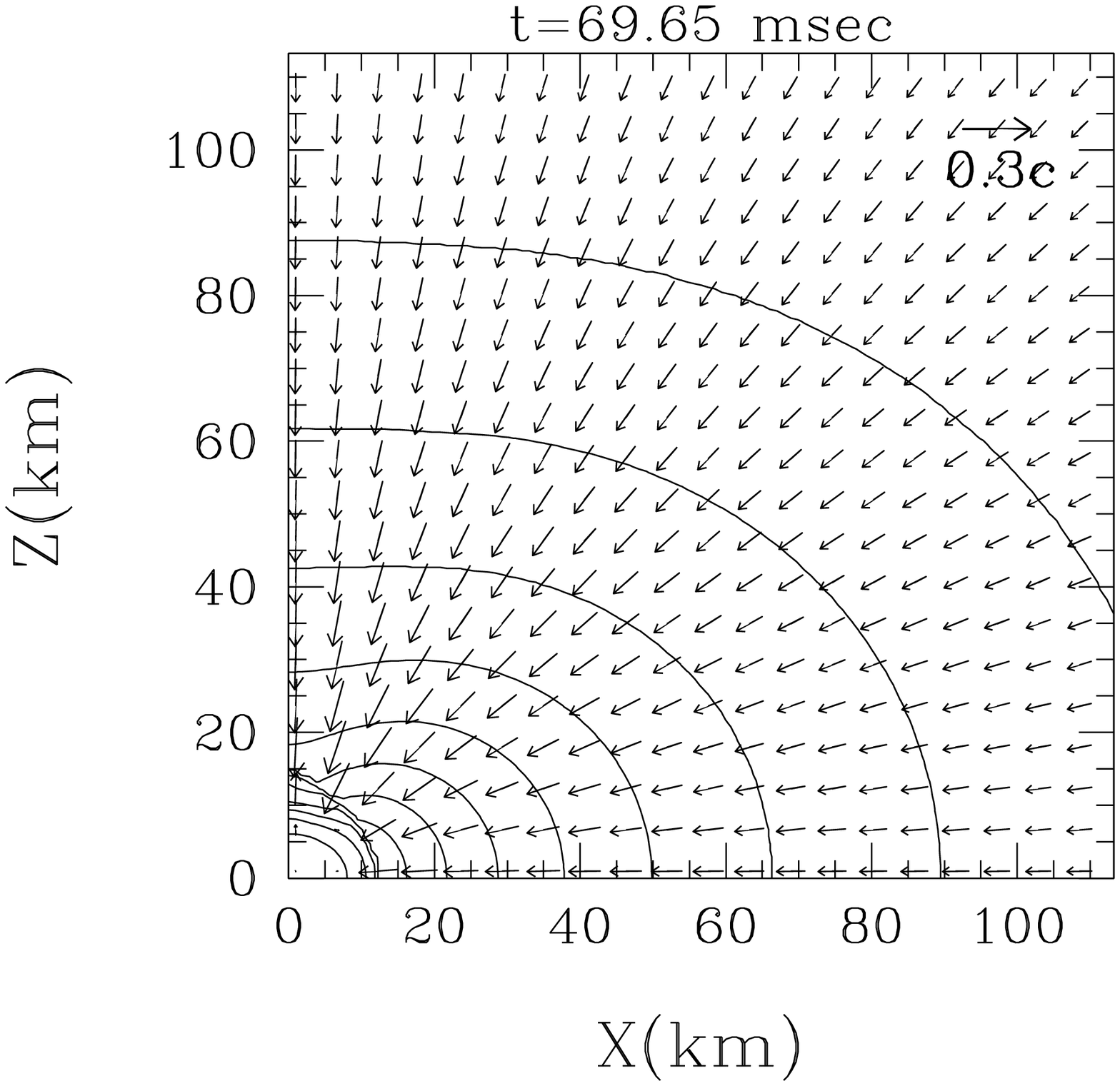}
\epsfxsize=2.2in
\leavevmode
\epsffile{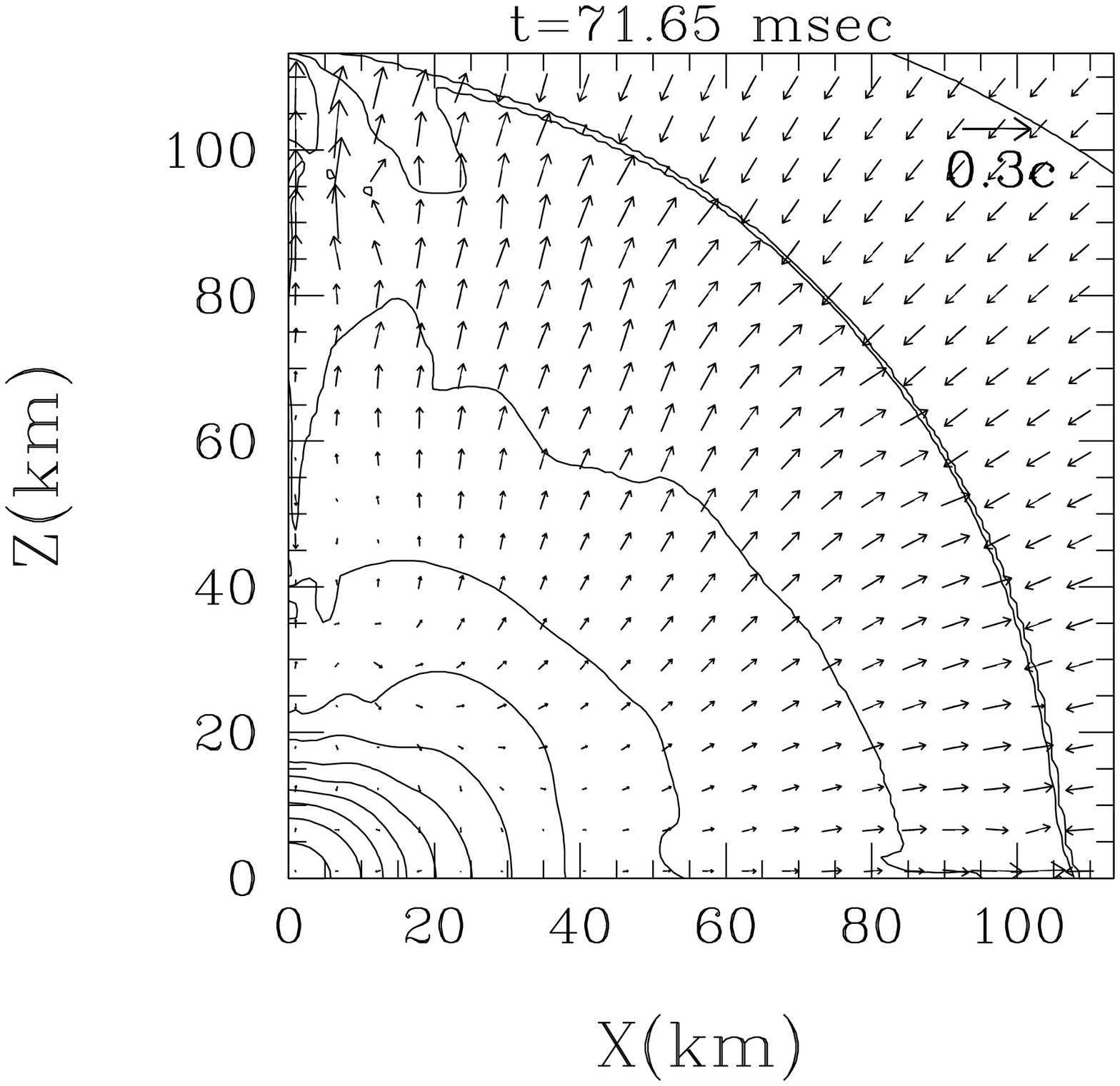}
\caption{Snapshots of the density contour curves of $\rho$ and
of the velocity field of $(v^x, v^z)$ at selected time slices
around which shocks are formed for model A1. 
The contour curves are drawn for
$\rho/\rho_{\rm nuc}=3\times 10^{-0.4j}$, 
with $j=0,1,2,\cdots,15$. 
\label{FIG5}
}
\end{center}
\end{figure}

\begin{figure}[htb]
%\vspace*{-4mm}
\begin{center}
\epsfxsize=2.2in
\leavevmode
\epsffile{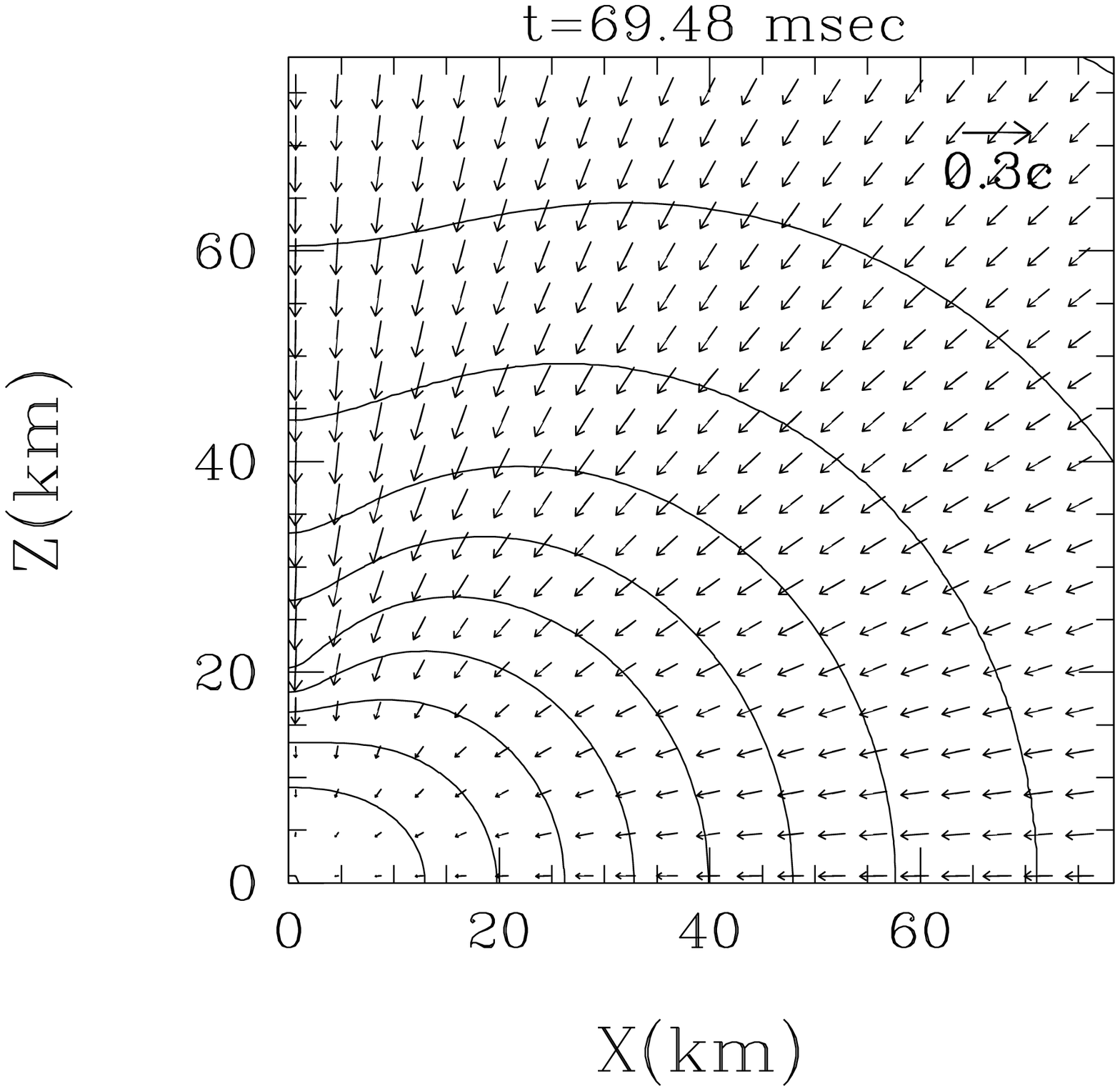}
\epsfxsize=2.2in
\leavevmode
\epsffile{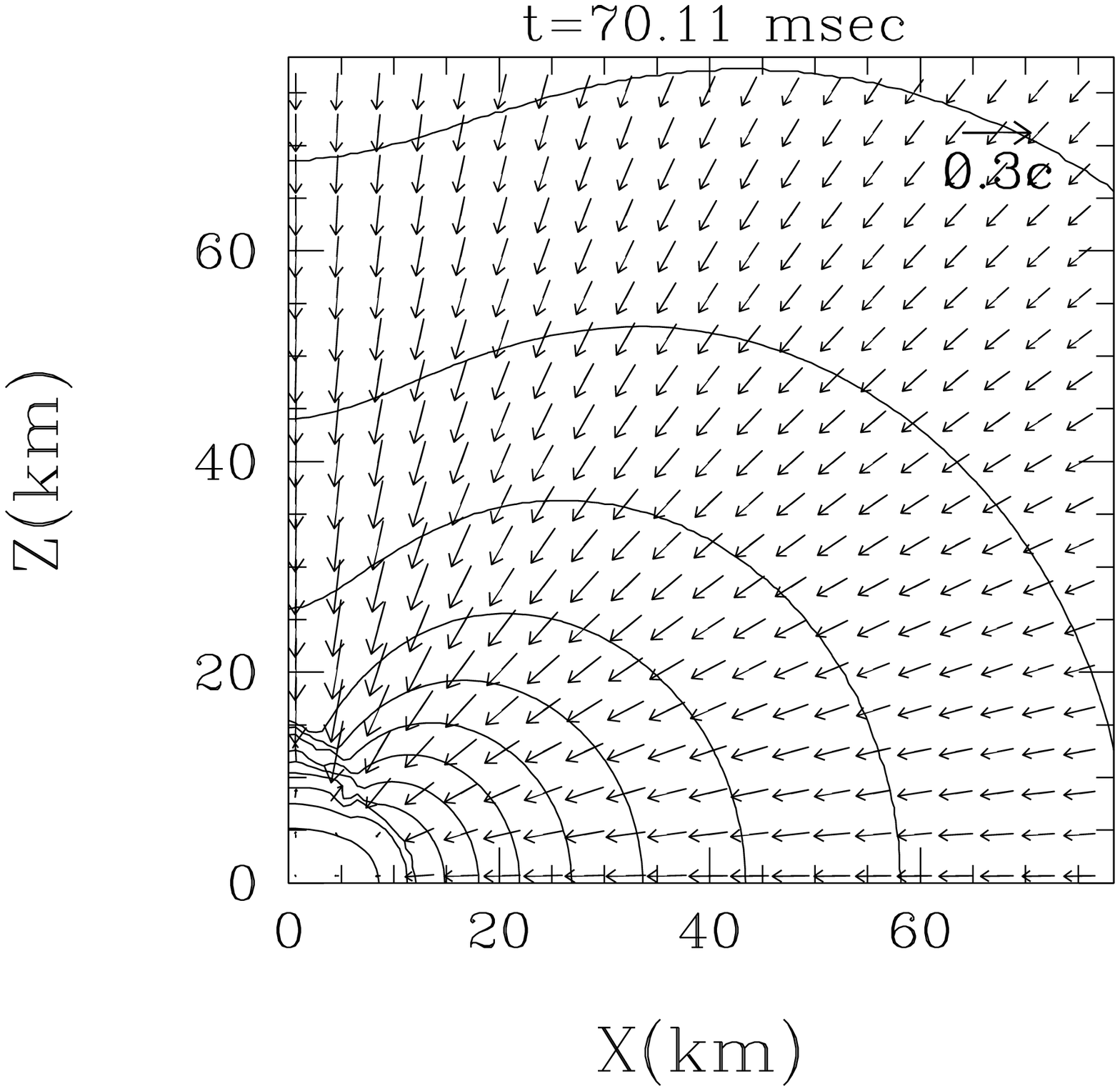}
\epsfxsize=2.2in
\leavevmode
\epsffile{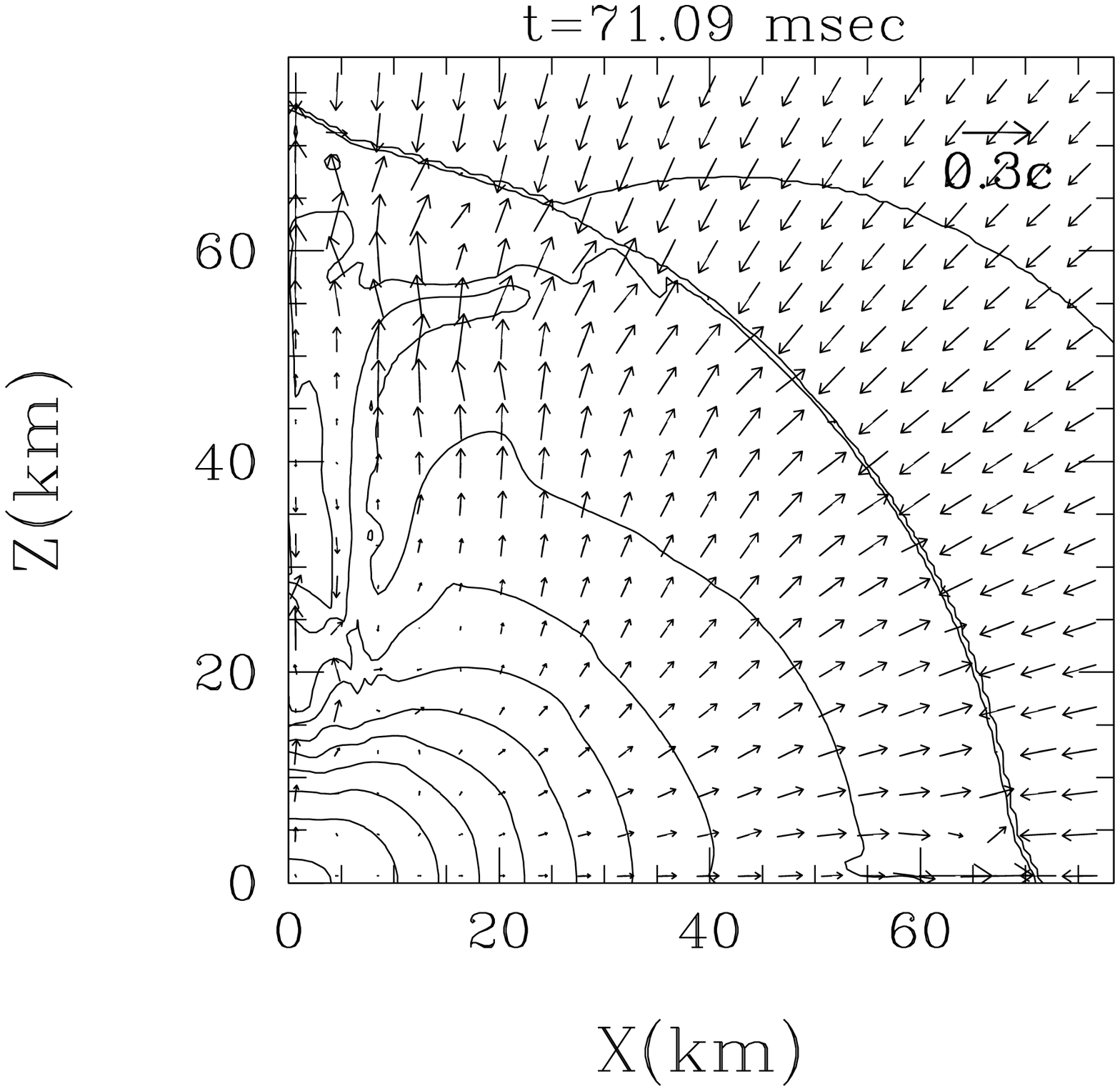}
\caption{The same as Fig. 5 but for model C1. 
The contour curves are drawn for $\rho/\rho_{\rm nuc}=3\times 10^{-0.4j}$, 
with $j=0,1,2,\cdots,15$. 
\label{FIG6}
}
\end{center}
\end{figure}

\begin{figure}[htb]
%\vspace*{-4mm}
\begin{center}
\epsfxsize=3.25in
\leavevmode
(a)\epsffile{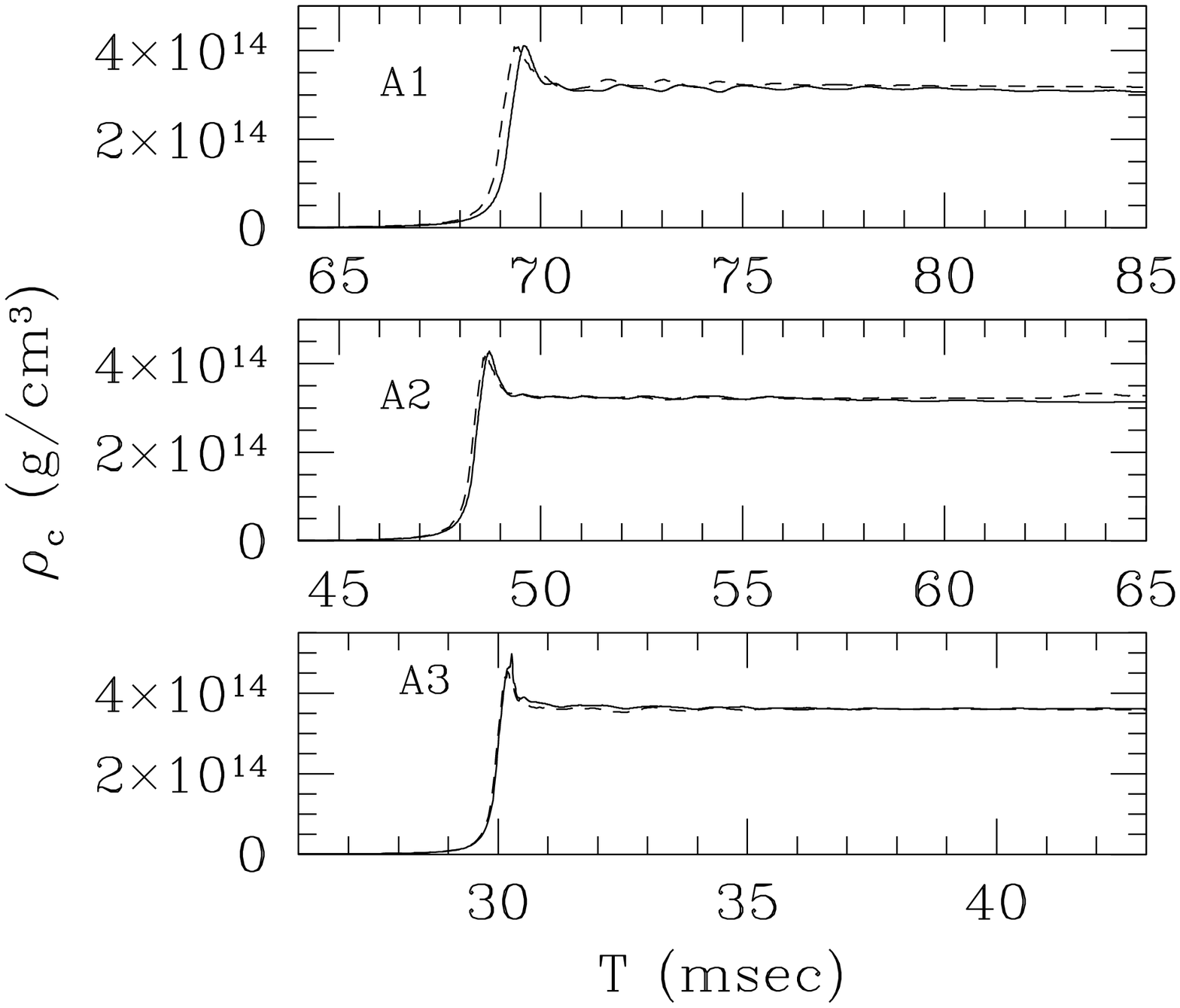}
\epsfxsize=3.25in
\leavevmode
(b)\epsffile{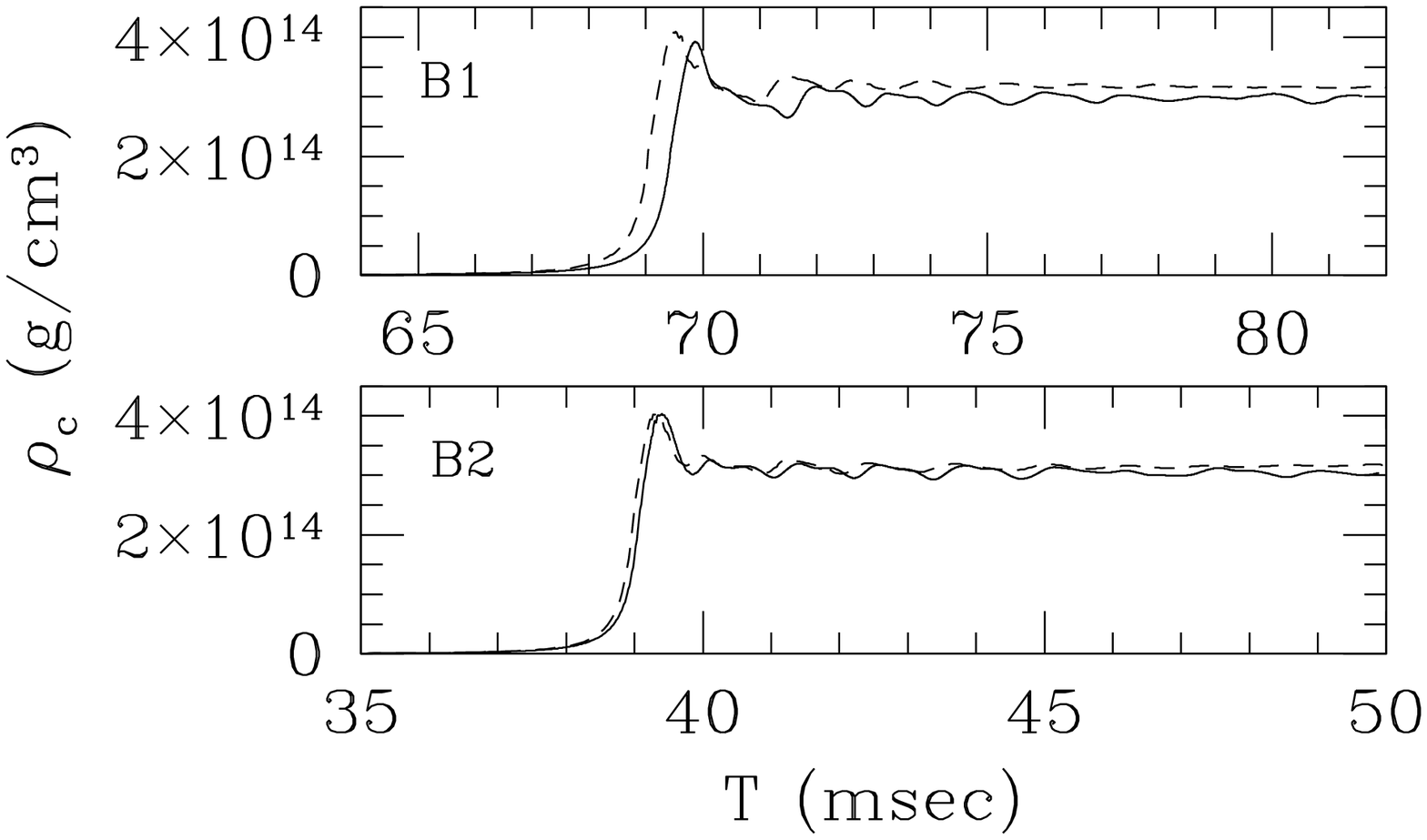}
%\vspace*{-4mm}
\caption{
Comparison between the evolution of the central density computed
in this paper (solid curves) and in Dimmelmeier et al. (dashed curves)
(a) for models A1--A3 and (b) for models B1 and B2.
%In (b), the results for models C1 and C2 are also plotted by the
%dotted curves. 
\label{FIG7}
}
\end{center}
\end{figure}

\subsubsection{Comparison with a previous work}

Here, we compare the numerical results 
for models A1, A2, A3, B1 and B2 with those 
for models A1B3G2, A1B3G3, A1B3G5, A3B2G2, and A3B2G4
in \cite{HD}, respectively. For these models, 
both groups adopt the almost identical initial conditions. 

Table IV shows the time at a maximum density achieved, the maximum density, 
and the maximum amplitude of gravitational waves for the numerical
results computed by two groups. In Fig. 7, we also compare the
evolution of the central density. It is found that the numerical results
by two groups agree within a small error both for models A and B. Only 
for model B1, the time at the maximum density achieved slightly 
disagrees with that for A3B2G2 by $\sim 0.3$ msec, but besides this
disagreement, the shape of $\rho_c$ as a function of time agrees well
each other even in this case. Recall that in \cite{HD}, the 
conformal flatness approximation to the Einstein equation
is adopted while our results are fully general relativistic. 
This indicates that the conformal flatness approximation is a 
good approximate formulation of general relativity for computing
axisymmetric rotating stellar core collapses to a neutron star. 

In a precise comparison, 
the following small systematic disagreements between two
results should be also addressed: 
(i) the maximum density achieved in our results is slightly larger 
for model A and slightly smaller for model B; 
(ii) the time at the maximum density is 
slightly delayed in our results, and this tendency is stronger
for the larger value of $\Gamma_1$ (i.e., for the longer infall time); 
(iii) for the larger value of $\Gamma_1$, 
the central density in the relaxed final stage is slightly smaller 
in our results. 

It is difficult to specify the particular reason for these
disagreements. There are several plausible candidates. 
First, computational settings are different between two groups.
In our simulation,
we adopted a uniform grid changing the grid spacing and
grid number, while in \cite{HD},
200 radial grid points with the logarithmic grid spacing were taken
throughout the simulation. 
In our case, the grid spacing is smaller than 0.5 km in the bounce and 
ring-down phases, although it is larger than 0.5 km in the infall phase. 
On the other hand, the minimum grid spacing is about 0.5 km in \cite{HD}
for all the phases. These differences may yield the disagreements.
Actually, we find that varying the grid resolution results in a
small change of the time at the maximum density achieved 
for models A1 and B1 [cf. Figs. 1(d) and 3(a)]. 
Secondly, the slicing condition is slightly different between two groups. 
In \cite{HD}, the maximal slicing condition, 
$K_k^{~k}=0$, was adopted, while in our numerical simulation, 
the condition is only approximately satisfied \cite{shibata}:
The equations $K_k^{~k}=0=\pa_t K_k^{~k}$ lead to an elliptic-type
equation for $\alpha$. In the exact maximal slicing condition,
this equation is iteratively solved until a convergence is achieved. 
In our case, we stop the iteration before the complete convergence
is achieved to save the computational time. Thus, $K_k^{~k} \approx 0$. 
This difference may result in a systematic deviation of the
coordinate time at the maximum density. 
Thirdly, the initial conditions adopted by two groups are 
not completely identical, since the equilibrium rotating
stars for the initial conditions are computed with different
numerical implementations. 
The values of $T/W$ and $\hat A$ may well have disagreement of
magnitude $\alt 1\%$. This may affect the subsequent numerical evolution
slightly. 

On the other hand, the difference of 
the adopted formulations for the gravitational field
is unlikely to be the reason for the disagreement.
This is because the deviation of the conformal metric $\tilde \gamma_{ij}$
from $\delta_{ij}$ is very small (typical absolute magnitude is 
of order $\sim 10^{-3}$ for each component) in our numerical results. 
Therefore, we infer that the magnitude of the systematic error due to
the conformally flatness approximation seems to be smaller than
that due to other reasons. 

\begin{table}[tb]
\begin{center}
\begin{tabular}{|c|c|c|c|} \hline
Model & $t_b$ & $\rho_{\rm max}~({\rm g/cm^3})$
& $(rh_+^{\quad})_{\rm max}$ (cm) \\ \hline
A1     & 69.5 & 4.12 & 561 \\
A1B3G2 & 69.5 & 4.02 & 469 \\ \hline
A2     & 48.7 & 4.28 & 215 \\
A1B3G3 & 48.6 & 4.23 & 180 \\ \hline
A3     & 30.3 & 4.98 & 32.7 \\
A1B3G5 & 30.2 & 4.55 & 33.9 \\ \hline
B1     & 69.8 & 3.93 & 731 \\
%C1     & 70.2 & 3.69 & 813 \\
A3B2G2 & 69.5 & 4.10 & 596 \\ \hline
B2     & 39.3 & 3.92 & 182 \\ 
%C2     & 39.6 & 3.76 & 248 \\
A3B2G4 & 39.3 & 4.05 & 141 \\ \hline
\end{tabular}
\caption{Comparison between the present (upper)
and previous numerical results by Dimmelmeier et al. (lower).
The time at bounce, the maximum density achieved, and the maximum
amplitude of gravitational waves are shown for two numerical results. 
}
\end{center}
\vspace{-5mm}
\end{table}

\begin{figure}[htb]
%\vspace*{-4mm}
\begin{center}
\epsfxsize=2.95in
\leavevmode
(a)\epsffile{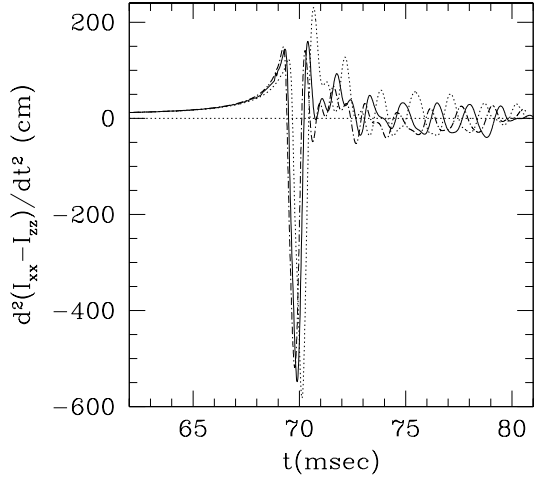}
\epsfxsize=2.95in
\leavevmode
(b)\epsffile{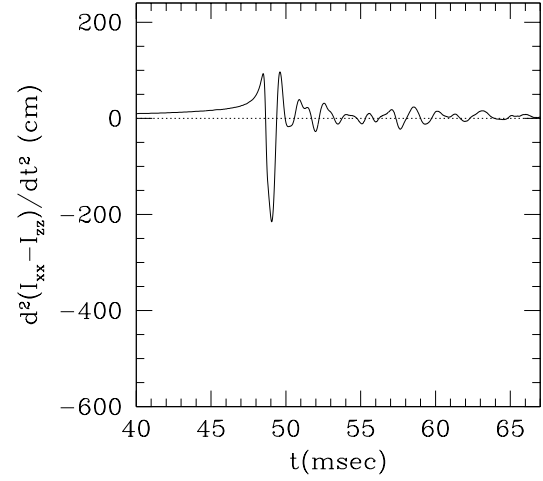}\\
\epsfxsize=2.95in
\leavevmode
(c)\epsffile{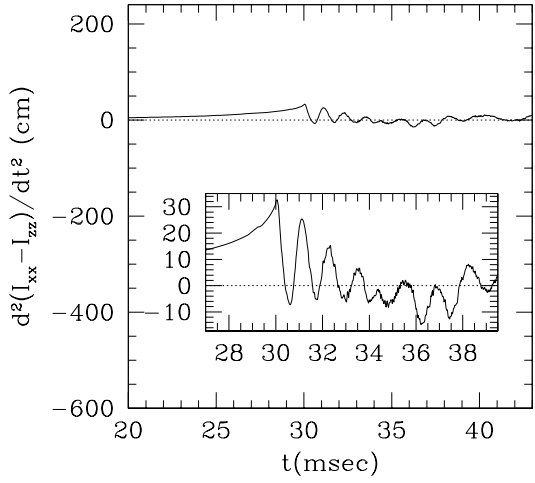}
\epsfxsize=2.95in
\leavevmode
(d)\epsffile{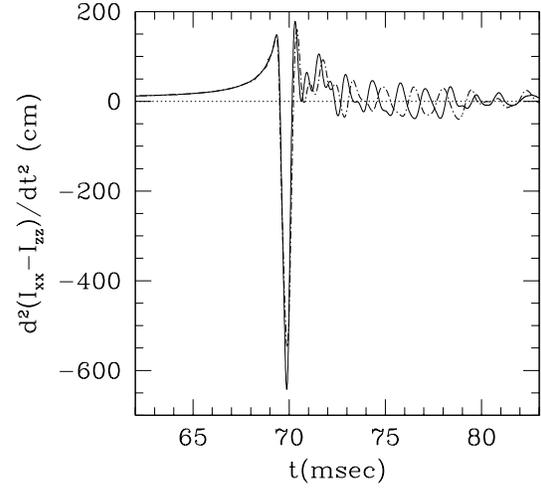}
\caption{Gravitational waveforms for model A; 
(a) Models A1 (solid curve), A4 (dotted-dashed curve),
and A5 (dotted curve); (b) Model A2; (c) Model A3; 
(d) Models A6 (solid curve) and A1 (dashed curves)
%%(d) Model A1 with high (solid curve) and low (dashed curves)
grid resolutions. 
\label{FIG8}
}
\end{center}
\end{figure}

\begin{figure}[htb]
%\vspace*{-4mm}
\begin{center}
\epsfxsize=2.95in
\leavevmode
(a)\epsffile{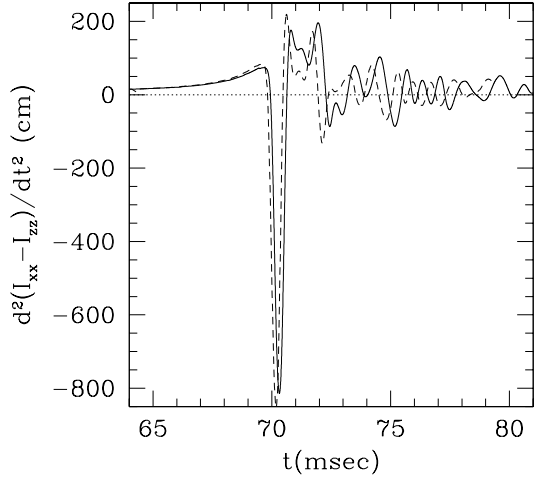}
\epsfxsize=2.95in
\leavevmode
(b)\epsffile{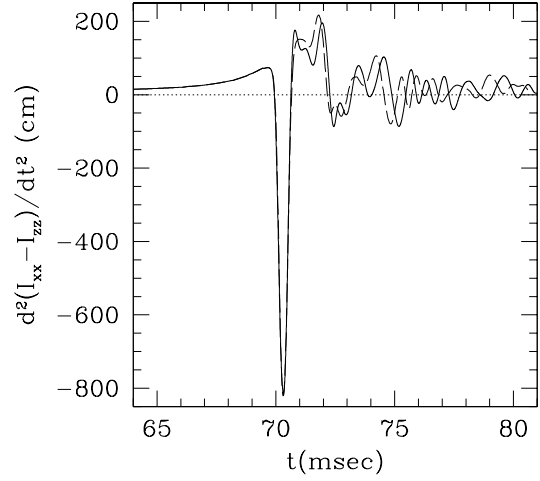}\\
\epsfxsize=2.95in
\leavevmode
(c)\epsffile{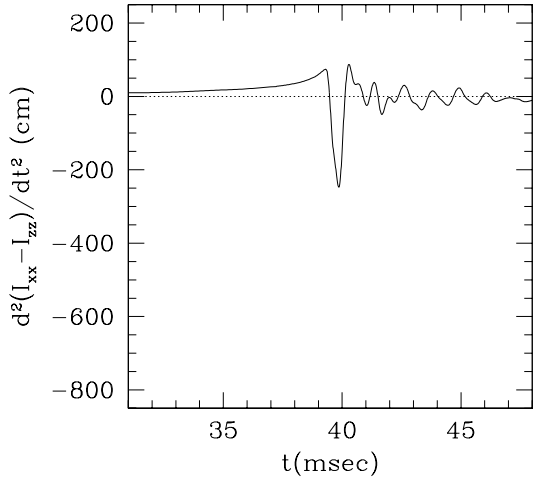}
\epsfxsize=2.95in
\leavevmode
(d)\epsffile{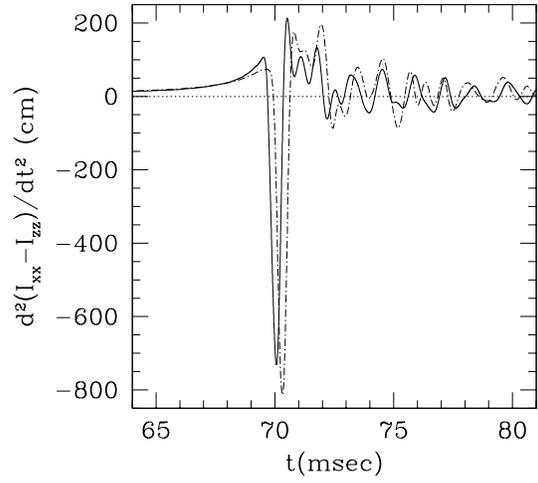}
\caption{Gravitational waveforms for model C; 
(a) Models C1 (solid curves) and C4 (dashed curves); 
(b) Models C1 (solid curves) and C3 (long-dashed curves); 
(c) Model C2; 
(d) Models B1 (solid curve) and C1 (dotted-dashed curve).
\label{FIG9}
}
\end{center}
\end{figure}

\begin{figure}[htb]
%\vspace*{-4mm}
\begin{center}
\epsfxsize=2.95in
\leavevmode
\epsffile{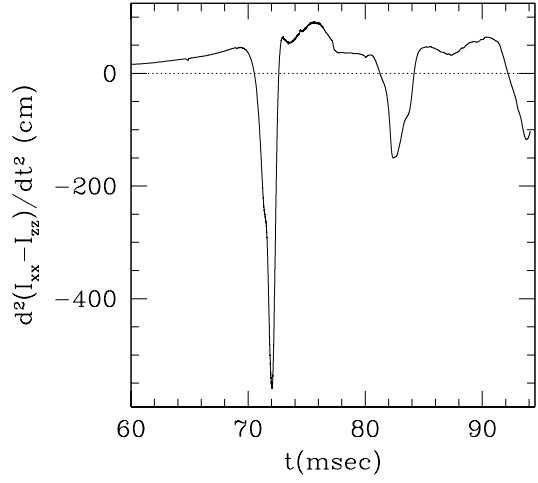}
\caption{Gravitational waveforms for model D. 
\label{FIG10}
}
\end{center}
\end{figure}

\begin{figure}[htb]
%\vspace*{-4mm}
\begin{center}
\epsfxsize=2.95in
\leavevmode
\epsffile{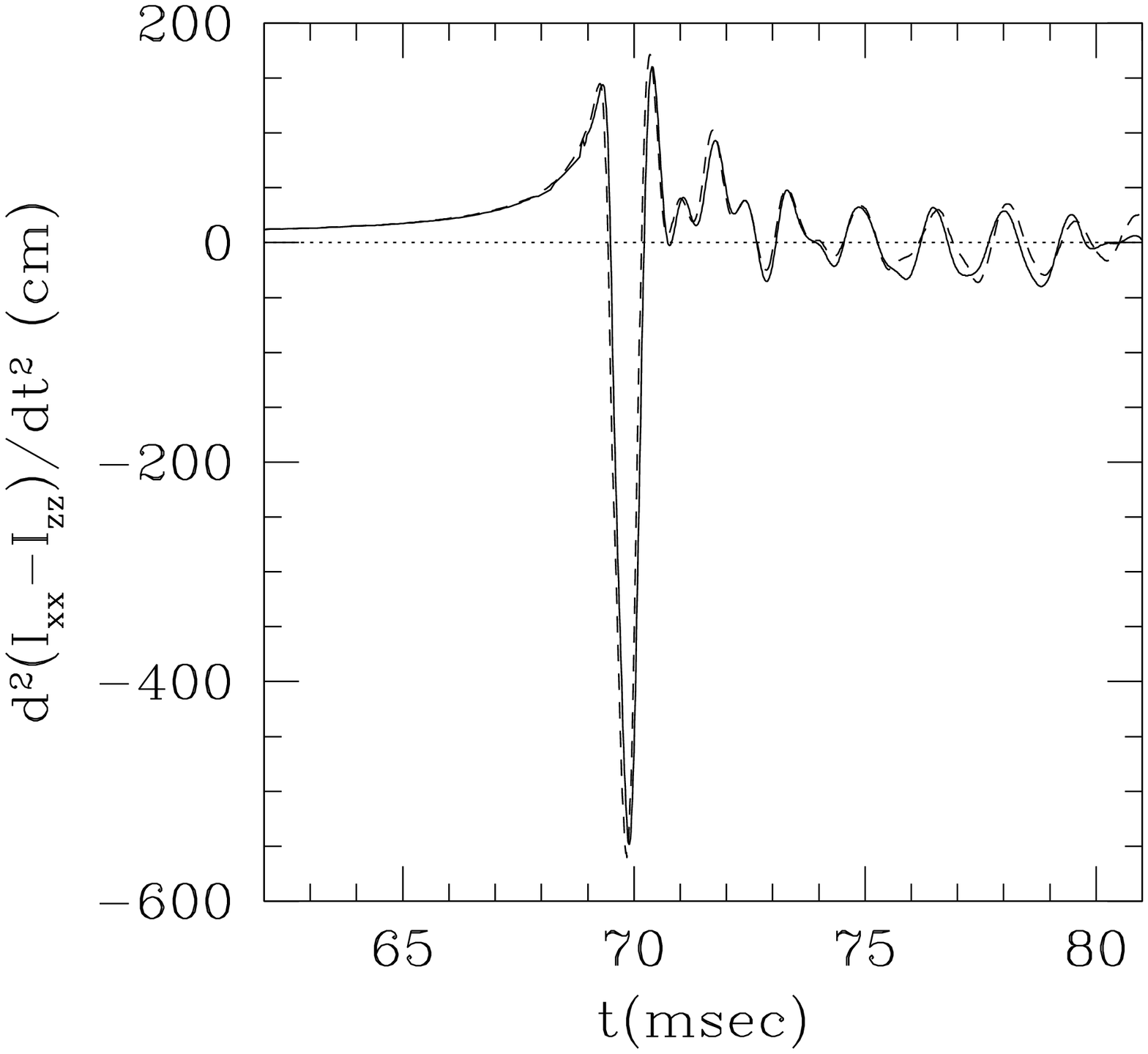}
\epsfxsize=2.95in
\leavevmode
\epsffile{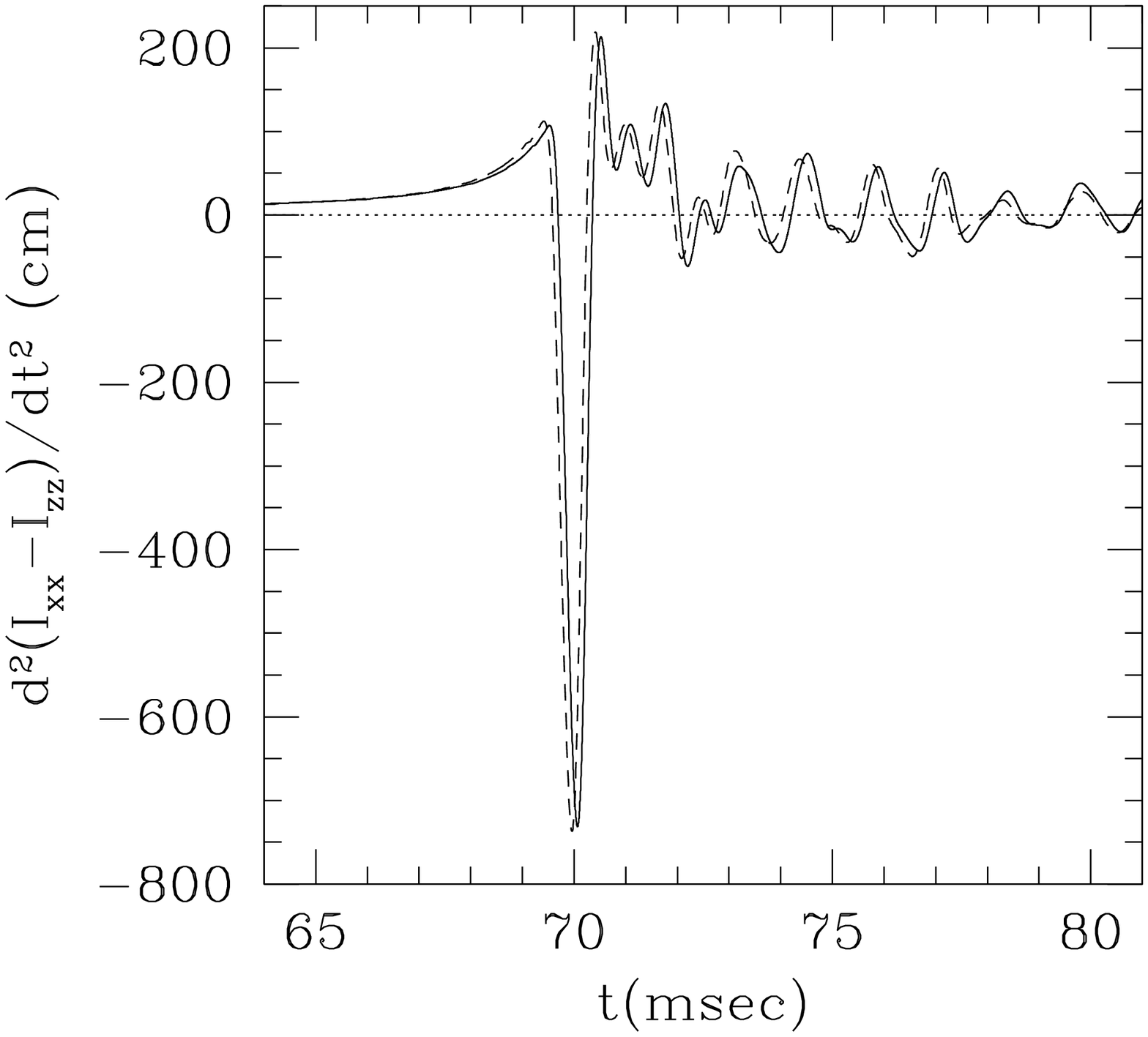}
\caption{
Gravitational waveforms (a) for model A1 and (b) for model B1 
with high (solid curve) and low grid resolutions (dashed curve). 
\label{FIG11}
}
\end{center}
\end{figure}

\subsection{Gravitational waveforms}
\subsubsection{General feature}

Gravitational waveforms are computed in terms of
the quadrupole formula described in Sec. II C. 
Since fully general relativistic simulations 
are performed, gravitational waves should be computed from 
the metric in a wave zone. 
However, we have found that it is not possible, 
since the amplitude is smaller than the numerical noise. 
An estimate by the quadrupole formula indicates that
the maximum amplitude of gravitational waves is 
smaller than $10^{-5}$ in the local wave zone for $r \sim \lambda$ 
where $\lambda$ denotes the wave length which is 
typically several hundred km. 

As illustrated in a previous paper \cite{SS}, approximate gravitational 
waveforms can be computed in terms of a quadrupole formula for 
highly relativistic, highly oscillating, and rapidly rotating 
neutron stars. In rotating stellar core collapses to a
neutron star, gravitational waves
are dominantly emitted during the bounce and ring-down phases.
Such gravitational waves are excited by 
oscillations of a formed protoneutron star. 
Thus, it is likely that the present approach can yield 
high-quality approximate gravitational waveforms besides
possible underestimation of the amplitude by $\sim 10\%$
due to absence of higher general relativistic corrections. 

Figures 8(a)--(d) show gravitational waveforms for model A 
with various sets of $\Gamma_1$, $\Gamma_2$, and $\Gamma_{\rm th}$. 
The waveforms for models A1, A2, A4, A5, and A6 are classified into 
type I according to Dimmelmeier et al. \cite{HD}. 
Properties of the type I gravitational waveforms 
can be summarized as follows: During the infall phase, a precursor 
whose amplitude and characteristic frequency increase monotonically
with time is emitted due to the infall and the flattening of 
the rotating core. The duration of the infall phase is $\agt 40$ msec 
and longer than a dynamical time scale defined at $t=0$ as 
$\rho_c^{-1/2} \sim 40$ msec. 
In the bounce phase, spiky burst waves are emitted 
for a short time scale $\sim 1$ msec, and the amplitude and the 
frequency of gravitational waves become maximum. In the ring-down phase, 
gravitational waves associated with several oscillation modes of a 
formed protoneutron star are emitted and its amplitude is gradually 
damped due to shock dissipation at the outer edge of the protoneutron star. 

For model A3 [cf. Fig. 8(c)] for which the simulation is
performed with a small value of $\Gamma_1$ as 1.28, the waveforms are 
qualitatively different from those for others: 
A sharp and distinguishable peak is not found at the bounce. 
Soon after the precursor emitted during the infall phase, 
the ring-down waveforms appear to be excited. 
An outstanding feature is that the amplitude in this case is 
much smaller than that for $\Gamma_1=1.31$ and 1.32 although the 
wave length is not significantly different from those for 
other models. According to \cite{HD}, 
this type of the waveforms is classified into type III.

In Fig. 8(a), the waveforms for models A1, A4, and A5 are presented.
For these models, we adopt $\Gamma_1=1.32$ and $\Gamma_2=2.5$, so that 
only the value of $\Gamma_{\rm th}$ is different. 
In the infall phase, the waveforms for three models are very similar.
This is natural because as long as the density is smaller
than $\rho_{\rm nuc}$, the magnitude of $P_{\rm th}$ is
much smaller than that of the cold part. Clear 
differences in the wave phase, wave length, and amplitude are
observed in the bounce and ring-down phases.
The reasons for them are explained as follows: 
%%For the smaller value of $\Gamma_{\rm th}$, the magnitude of
%%$P_{\rm th}$ is smaller and, thus, the shock heating is weaker. 
The smaller magnitude of $P_{\rm th}$ results in the slightly 
shorter infall time as reflected in the time at which 
the amplitude becomes maximum. As a consequence, the difference of the 
wave phase is yielded. Stronger shock heating, which 
generates larger thermal energy, also results in smaller compactness of the 
formed protoneutron stars. This leads to the results that
for the larger value of $\Gamma_{\rm th}$, 
the gravitational wave length, which in general increases with 
the stellar radius for a given mass, becomes longer, 
and the amplitude, which is larger for stronger shock heating, 
is larger.

Slight change of the value of $\Gamma_1$, which
determines the dynamics of the infall phase, 
significantly modifies gravitational waveforms. 
Comparison among Figs. 8(a)--(c) clarifies 
that with the decrease of the value of $\Gamma_1$,
the amplitude of gravitational waves decreases systematically. 
The reason for this is explained as follows: 
For the smaller value of $\Gamma_1$, 
the central region collapses more rapidly than the outer region does. 
This results in a smaller core mass at the bounce
for the smaller value of $\Gamma_1$.
The amplitude of gravitational waves increases with the increase of 
the core mass for a fixed value of the density and, therefore,
it is smaller for the smaller value of $\Gamma_1$. 
%%%This property agrees with what is found in \cite{HD}. 

In Fig. 8(d), we compare the waveforms of different values of 
$\Gamma_2$ with fixed values of $\Gamma_1$ and $\Gamma_{\rm th}$.
It is found that the difference of the waveforms between two models appears
only in the bounce and ring-down phases. This is natural because the
value of $\Gamma_2$ does not affect the infall phase and mainly 
determines the equations of state and the radius (or compactness) of 
the formed protoneutron stars. Recall that the smaller value of
$\Gamma_2$ results in the larger compactness of the protoneutron star. 
This fact is reflected in slightly shorter wave length 
and larger amplitude of gravitational waves
in the ring-down phase for the smaller value of $\Gamma_2$. 

Figure 9 displays gravitational waveforms for model C. 
As in the case of model A, the waveforms are divided into three parts
(precursor, spike, and ring-down), but the qualitative 
feature of the ring-down waveforms between models A and C is different. 
For example, compare the waveforms for models A1 and C1 for which 
the values of $\Gamma_1$, $\Gamma_2$, and $\Gamma_{\rm th}$ are identical. 
For model A1, the waveforms are modulated only in the
early ring-down phase (e.g., for $t\sim 70$--73 msec). 
In the late ring-down phase (e.g., for $t \agt 73$ msec for model A1), 
they are fairly periodic and appear to be composed mainly of one or two
eigen oscillation modes of the formed protoneutron star. 
On the other hand, for model C1, the waveforms are not very periodic 
and highly modulated throughout the ring-down phase.
In this case, several eigen modes of a formed protoneutron star
appear to constitute gravitational waveforms. 
Such modulated waveforms are likely to be due to the fact that 
the formed protoneutron star is rapidly and differentially rotating and 
the oscillation modes are excited in a complicated manner at the bounce. 

In Fig. 9(a), we compare the waveforms of different values of 
$\Gamma_{\rm th}$ with fixed values of $\Gamma_1$ and $\Gamma_2$.
As in Fig. 8(a), for the smaller value of $\Gamma_{\rm th}$, 
the maximum amplitude is reached at an earlier time, 
the wave length during the bounce and ring-down phases is 
longer, and the amplitude is smaller. 
These are universal features independent of the initial
rotational velocity profiles. 
However, in contrast to Fig. 8(a), the waveforms in the ring-down
phase for models C1 and C4 are not very similar.
Thus, small change of $\Gamma_{\rm th}$ from 1.35 to 1.5
significantly modifies the ring-down waveform in the case
of differentially rotating initial velocity profiles. 

In Fig. 9(b), we compare the waveforms of different values of
$\Gamma_2$ with fixed values of $\Gamma_1$ and $\Gamma_{\rm th}$.
In contrast to Fig. 8(d),
the maximum amplitude of gravitational waves is nearly identical
for two models. 
This suggests that in halting the infall, 
the centrifugal forces may play an important role to hide the effects
of the difference in the value of $\Gamma_2$. 
The difference of the ring-down waveforms between two models
is qualitatively the same as that found in Fig. 8(d): 
For the smaller value of $\Gamma_2$, 
the wave length and amplitude of gravitational waves 
in the ring-down phase are 
slightly shorter and larger, respectively.

In Fig. 9(c), the waveform for model C2 is displayed. This should 
be compared with the solid curve in Fig. 9(a) [or 9(b)] for model C1
of a different value of $\Gamma_1$. 
Comparison between two waveforms shows that with the decrease of 
the value of $\Gamma_1$, the wave amplitude at the bounce
and ring-down phases decreases. This property agrees with 
that found for model A and is likely to be
independent of the initial rotational velocity profiles. 

To see the effect of the slight change of differential rotation
parameter $\hat A$, 
we compare the waveforms of models B1 (solid curve) and
C1 (dotted-dashed curve) in Fig. 9(d).
Two waveforms are qualitatively similar, but 
for model C1 the amplitude is larger and the modulation of the
amplitude is induced more. 
This illustrates that with a slight modification of the initial rotational 
velocity profile, the resulting gravitational waveforms are modified
significantly. 

Figure 10 shows the gravitational waveform for model D. 
In this model, the collapse does not lead to a quasistationary 
protoneutron star of $\rho_c > \rho_{\rm nuc}$.
Instead, a quasiradially oscillating star of 
subnuclear density is formed and, therefore, quasiperiodic waves 
of a long period $\sim 10$ msec are emitted. 
According to \cite{HD}, this is classified into the type II waveform.

Convergence of the numerical results appears to be achieved. 
In Fig. 11, we display the numerical results with high and low grid
resolutions for models A1 and B1. 
The grid spacing in the low grid resolution 
is about 5/3 as large as that in the high case. 
It is found that the computed gravitational waveforms depend 
only weakly on the grid resolution in our choice of the grid spacing. 
We conclude that the grid resolution we choose in this work is fine 
enough to compute convergent gravitational waveforms. 

\subsubsection{Comparison with a previous work}

Here, we compare the gravitational waveforms computed in this paper 
with those in \cite{HD} for models A1, A2, A3, B1, and B2. 
Figures 12 and 13 show the gravitational waveforms 
computed by us (solid curves) and by Dimmelmeier et al. \cite{HD}
(dashed curves). It is found that 
the waveforms in the infall phase agree very well each other. 
In the bounce phase, on the other hand, the amplitude of our results is 
larger than that in \cite{HD} by $\sim 20\%$ for models A1, A2, B1,
and B2, although they still agree qualitatively.
The disagreement is outstanding in the ring-down phase. 
The amplitudes of gravitational waves in the ring-down phase
for models A1, A2, B1, and B2 are larger than those in \cite{HD} 
by a factor of $\sim 2$. Moreover, in our results, 
the oscillations with a nearly constant amplitude continue 
for several oscillation periods ($\agt 10$ msec). This is not the case 
in the results of \cite{HD} in which 
the amplitude is damped within several msec.

\begin{figure}[htb]
\vspace*{-4mm}
\begin{center}
\epsfxsize=2.9in
\leavevmode
(a)\epsffile{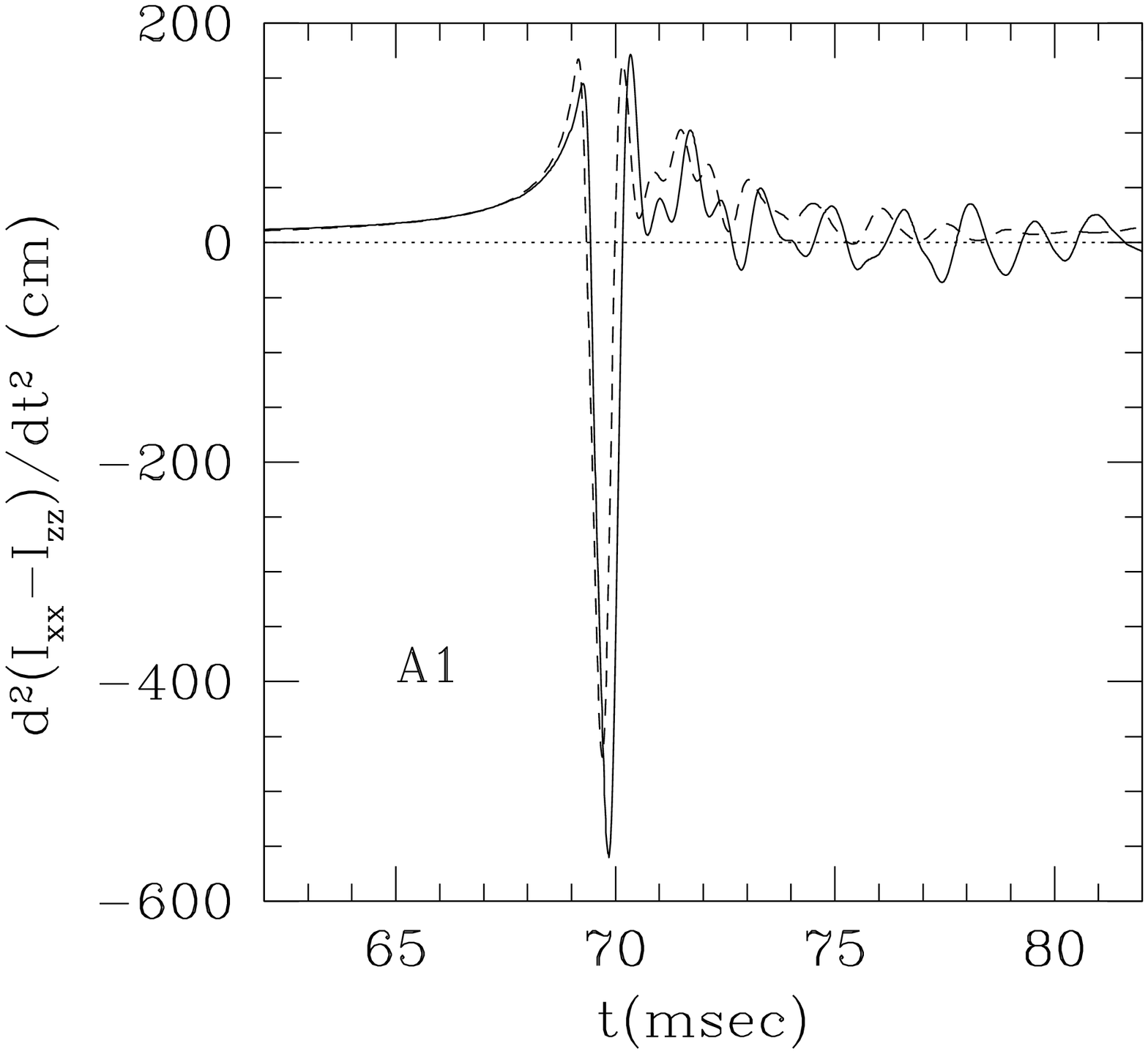}
\epsfxsize=2.9in
\leavevmode
(b)\epsffile{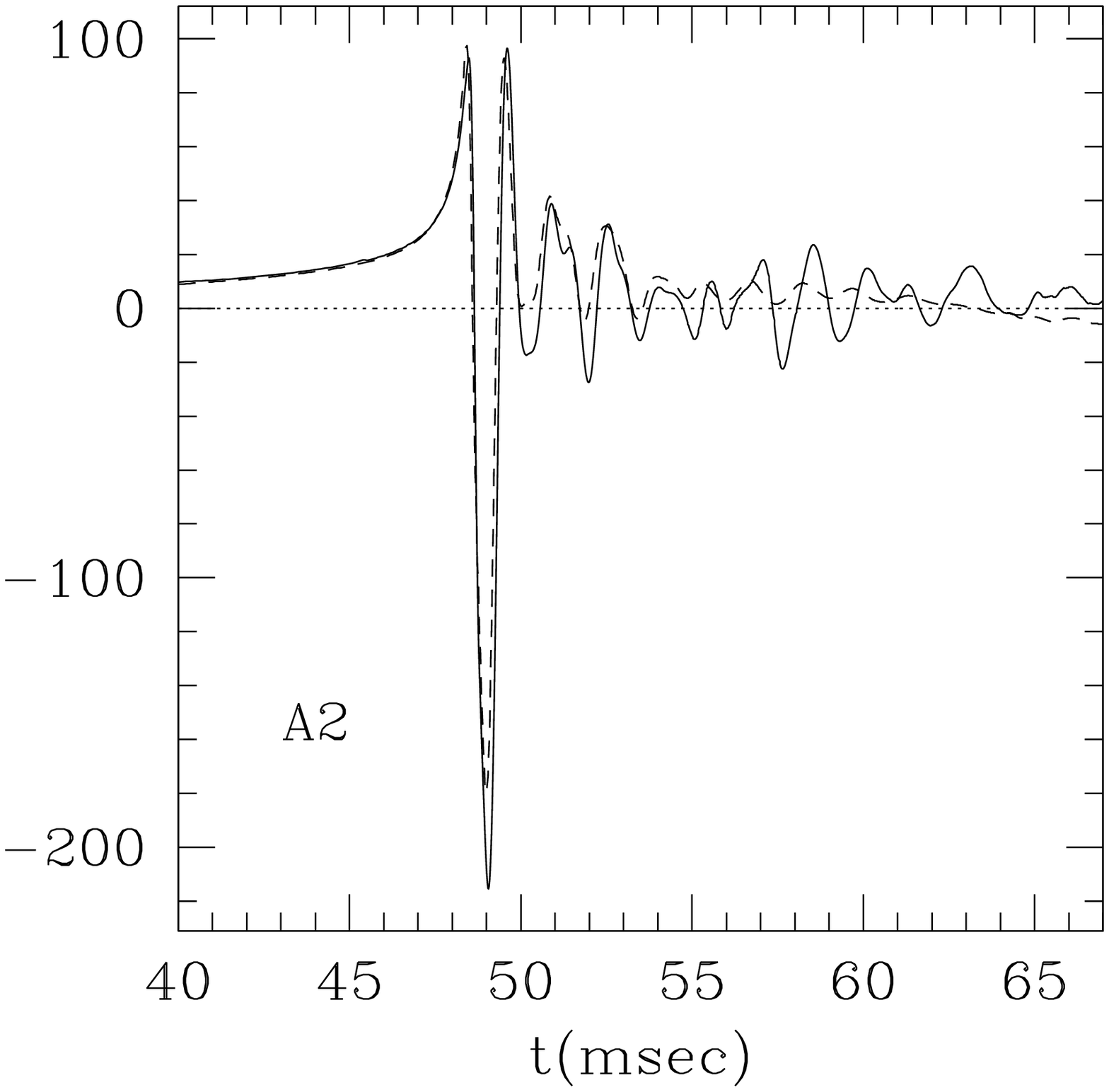} \\
\epsfxsize=2.9in
\leavevmode
(c)\epsffile{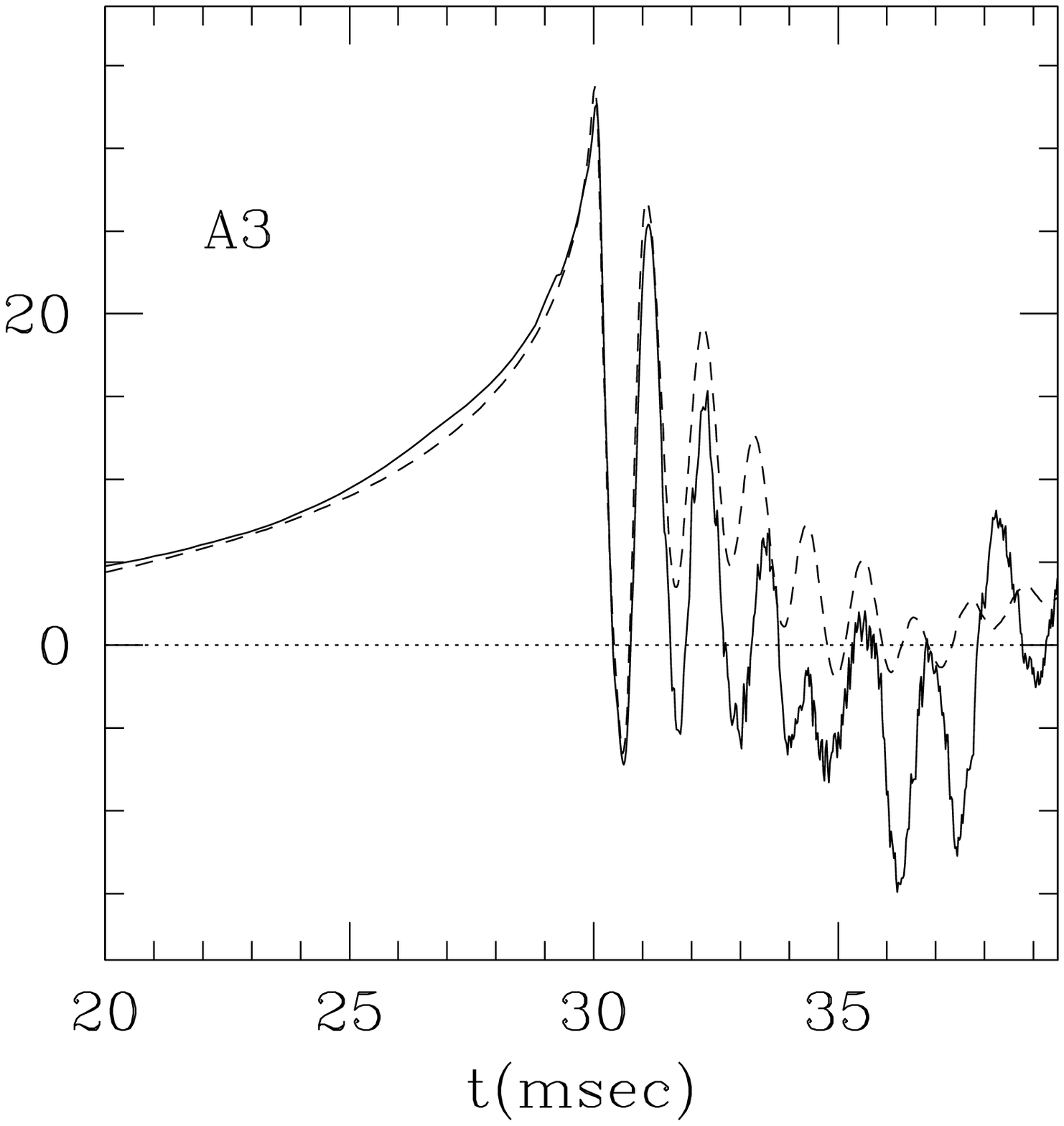}
\caption{Comparison between gravitational waveforms computed 
in this paper (solid curves) and in Dimmelmeier et al. (dashed
curves) for models A1--A3 [(a)--(c)]. 
\label{FIG12}
}
\end{center}
\end{figure}

\begin{figure}[htb]
\vspace*{-4mm}
\begin{center}
\epsfxsize=2.9in
\leavevmode
(a)\epsffile{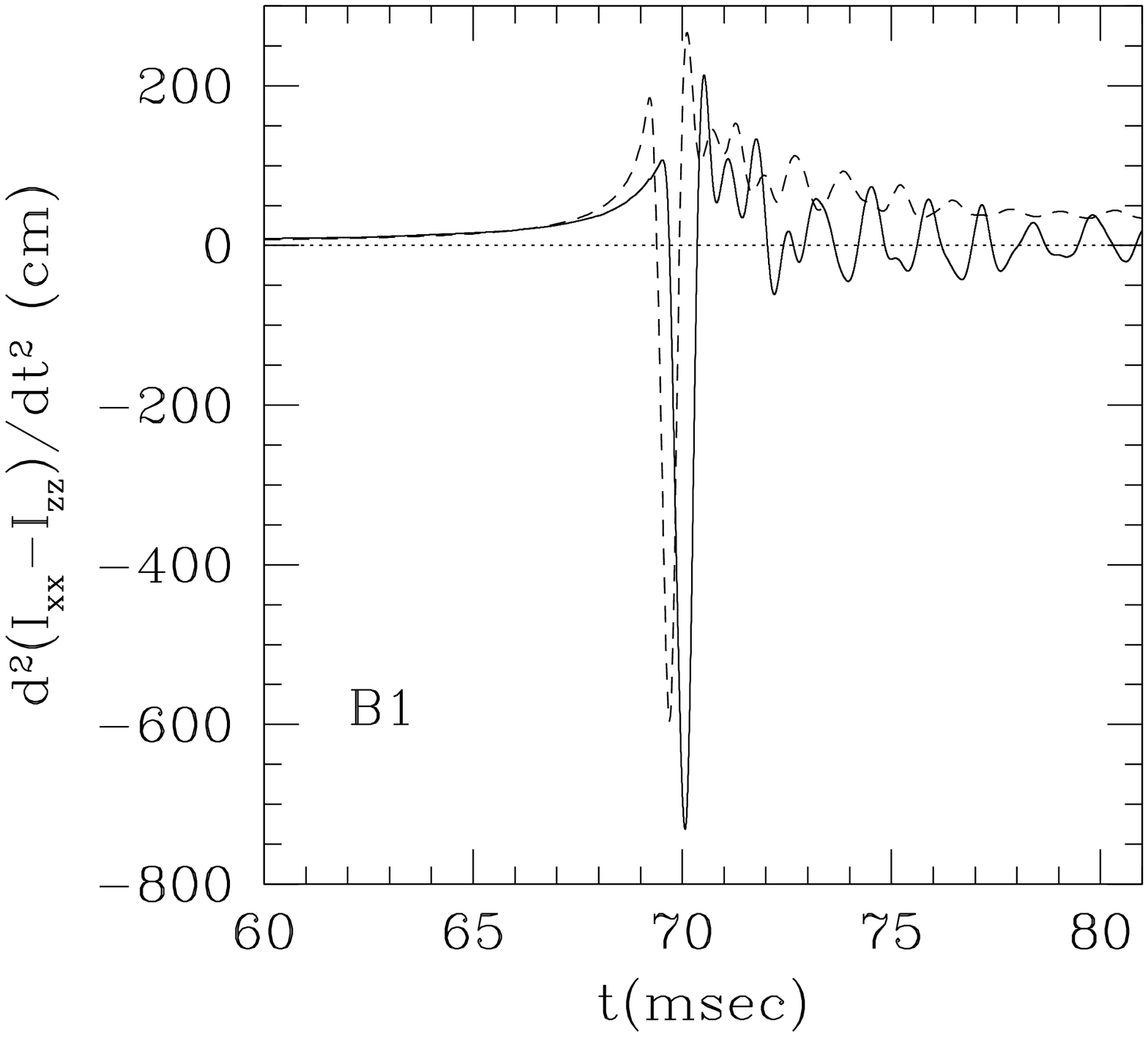}
\epsfxsize=2.9in
\leavevmode
(b)\epsffile{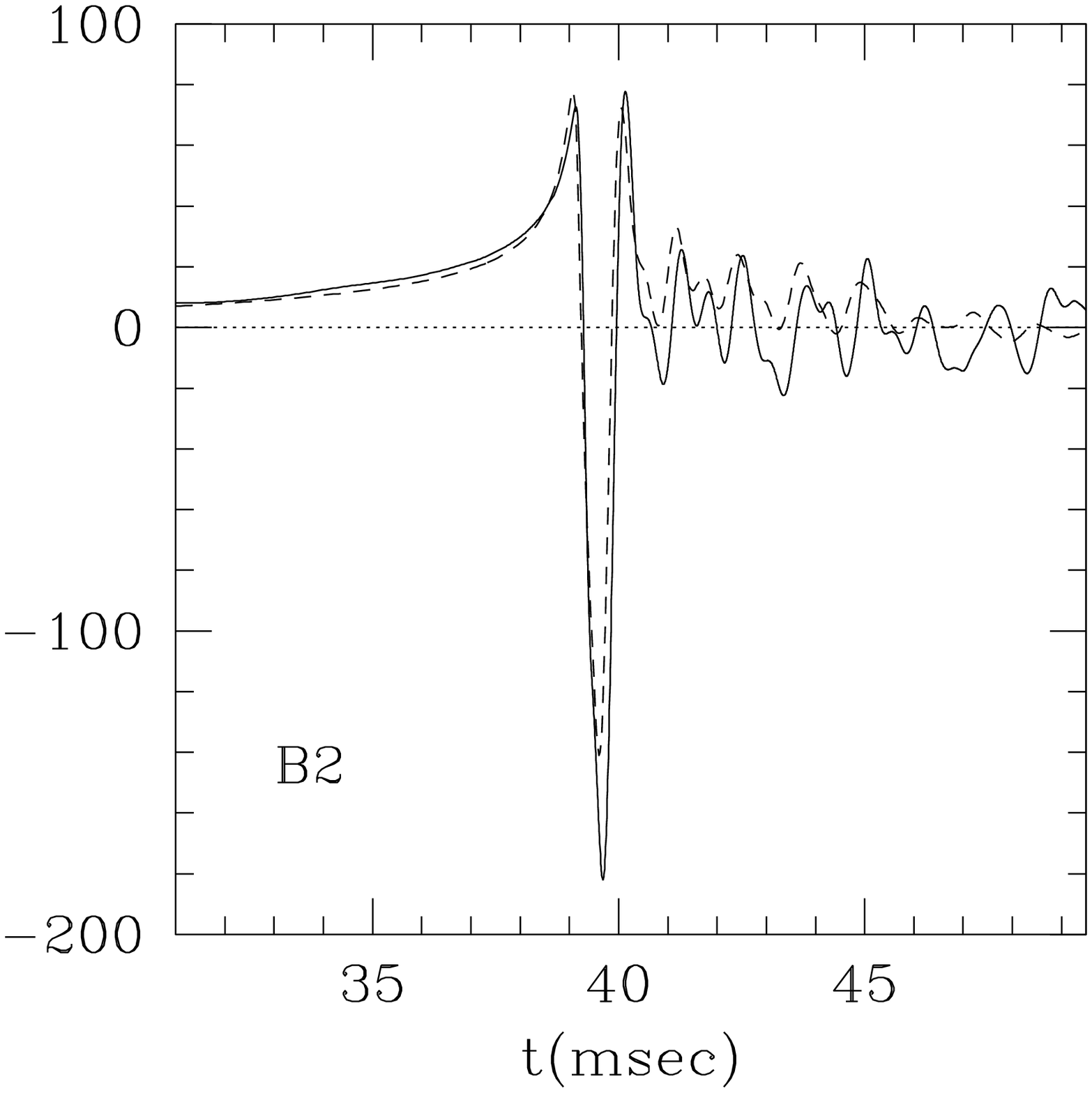}
\caption{The same as Fig. 12 but for models B1 and B2 [(a) and (b)]. 
\label{FIG13}
}
\end{center}
\end{figure}

This would be partly due to the difference in the grid 
resolution or in the slicing condition
adopted by two groups as mentioned in the previous section.
However, the main reason is likely that quadrupole formulas adopted 
by two groups are not identical. In the quadrupole formula we adopt, 
a quadrupole moment is simply defined using 
a weighted rest-mass density $\rho_*$ and then 
the second time derivative is taken with no approximation. 
In \cite{HD}, on the other hand, the quadrupole moment is defined 
using $\rho$ and, in addition, when taking the second time derivative, 
they discard higher relativistic terms keeping only the lowest order 
post-Newtonian terms. 

This disagreement raises a question: what is a good quadrupole
formula in general relativistic simulations ? 
An excellent quadrupole formula should yield a high-quality
approximate waveform for the true one computed from the metric in
the wave zone. Thus, to answer the question, it is necessary 
to compare the gravitational waveforms computed by a quadrupole formula 
with those extracted from the metric. In \cite{SS}, 
we calibrated the waveforms by performing simulations for highly
relativistic, highly oscillating, and rapidly rotating neutron stars
with $M/R \sim 0.2$ and $v/c \sim 0.3$ and 
found that our quadrupole formula yields well-approximated waveforms; 
the wave phases agree well with those computed from the metric 
and the wave amplitude is computed within an error of magnitude 
of $O(M/R)$ or $O(v^2/c^2)$. We believe that the waveforms presented
in this paper are well-approximated ones in phase and within $\sim 10\%$
error in amplitude. On the other hand, the quadrupole formula adopted 
in \cite{HD} has not been calibrated, since 
they adopted the conformal flatness approximation in which
gravitational waves cannot be extracted from the metric.
Thus, it is not clear how good their quadrupole formula is.
Since the amplitudes computed by our quadrupole formula are 
underestimated by $\sim 10\%$ and
the amplitudes computed in \cite{HD} are smaller than ours,
gravitational waveforms presented in \cite{HD} may contain an error 
of magnitude more than 10--20\%.

\section{Summary}

We performed axisymmetric numerical simulations of rotating 
stellar core collapses to a neutron star in full general relativity, 
paying particular attention to gravitational waveforms and 
to comparison of our results with previous results \cite{HD}. 
The Einstein field equations are solved in the Cartesian coordinates 
imposing an axisymmetric condition by the Cartoon method \cite{alcu}. 
The hydrodynamic equations are solved in the cylindrical coordinates 
(with the Cartesian coordinates restricting on the $y=0$ plane) 
using a high-resolution shock-capturing scheme 
with the maximum grid size $(2500,2500)$. A parametric equation of 
state is adopted to model collapsing stellar cores and formed 
protoneutron stars following Dimmelmeier et al. \cite{HD}. 
Gravitational waveforms are computed using a quadrupole formula
proposed in \cite{SS}. 

We choose moderately rapidly rotating stars as 
the initial conditions for which the value of $T/W$ is between 0.005 and 
$0.01$. Simulations are performed changing three 
parameters ($\Gamma_1$, $\Gamma_2$, and $\Gamma_{\rm th}$)
which characterize the equation of state. 
The dynamics of the collapse depends on the three parameters as 
well as $T/W$ and $\hat A$ of the initial condition. 
The dependence of the evolution of the system and gravitational 
waveforms on these five parameters is studied. 
The value of $\Gamma_1$ mainly determines 
the duration of the infall phase and the coherence of 
the early phase of the collapse.
For the smaller value of $\Gamma_1$, the infall time becomes shorter 
and the collapse is accelerated more in the central region. 
This results in that the core mass at the bounce is smaller 
and that the magnitude of $\Phi_c$ 
(which may be regarded as the depth of the gravitational potential) 
at the bounce is smaller for the smaller value of $\Gamma_1$.
The amplitude of gravitational waves becomes also smaller for the
smaller value of $\Gamma_1$. 

The value of $\Gamma_2$ determines the equation of state for formed
protoneutron stars. Thus, it does not affect the evolution during the 
infall phase. It determines the 
final value of the central density of the formed protoneutron star
and the gravitational waveforms emitted during the ring-down phase
in which the eigen oscillation modes of the protoneutron stars are
excited. The value of $\Gamma_{\rm th}$ determines the strength of 
shock waves. We choose this value as 1.35, 1.5, and 5/3
extending the work by Dimmelmeier et al. \cite{HD}. It is found that 
for the smaller value of this parameter, the 
shock heating becomes weaker and the amplitude of 
gravitational waves smaller. 

The values of $T/W$ and $\hat A$ play a significant role in determining 
the dynamics of collapse and the corresponding gravitational
waveforms in particular in the bounce and ring-down phases.
For the rigidly rotating case ($\hat A \rightarrow \infty$),
the maximum value of $T/W$ is $\approx 0.009$ which we choose
in this paper. Even in this maximum case,
the collapse leads to a neutron star irrespective of the values
of $\Gamma_1$, $\Gamma_2$, and $\Gamma_{\rm th}$.
This indicates that for rigidly rotating initial conditions,
the neutron stars are formed soon after the collapse,
irrespective of the angular velocity of the initial condition,
with our choice of the equations of state. 
For the differentially rotating case with $\hat A = 1/4$, 
the collapse does not lead to a neutron star but an oscillating star
of subnuclear density is formed for $T/W \agt 0.01$ since the 
centrifugal force is strong enough near the rotational axis. 
As shown in \cite{HD}, more rapidly rotating initial conditions 
with $T/W \gg 0.01$ may be constructed. For such high values of
$T/W$, a neutron star will not be formed soon after the collapse. 

With a slight change of $\hat A$ from 0.25 to 0.32 for the
initial condition, the angular velocity at the rotational axis
is changed by a large factor even if $T/W$ is approximately
identical. As a result of this change, 
the subsequent evolution of the collapse and gravitational waveforms
in the bounce and ring-down phases are modified significantly. 
This implies that the dynamics rotating stellar core collapses and the 
corresponding gravitational waveforms are sensitive 
not only to the equation of state but also to the initial
angular velocity profile. 

%%This indicates that to prepare templates
%%of gravitational waves, it is required to know a 
%%plausible initial rotational velocity profile of the 
%%stellar core if it exists. 

Several simulations are performed setting the same initial conditions 
as those adopted in \cite{HD}. It is found that the dynamics of the
collapse and the bounce for such initial conditions 
are very similar to those found in \cite{HD}, 
in which an approximate general relativistic gravity
(the conformal flatness approximation) is assumed.
This indicates that such approximate relativistic formulation 
is appropriate for computing axisymmetric rotating stellar core collapses
and the subsequent formation of protoneutron stars.
(Note that this is the conclusion for the formation of
neutron stars. This may not be the case for the black hole formation.) 

Gravitational waveforms are compared with a previous result \cite{HD}.
It is found that the waveforms are qualitatively in good agreement but 
not quantitatively with those in \cite{HD}. 
Either of two plausible elements could explain this disagreement. 
One is that the grid resolution and computational setting 
are different between two groups. 
This could modify the waveforms slightly. However, 
the main reason seems to be that the adopted quadrupole formulas by two 
groups are different. As mentioned in the previous section, 
there is no unique definition of the quadrupole formula 
for dynamical spacetimes in general relativity. 
This implies that when one attempts to use 
a quadrupole formula in a relativistic 
simulation, one needs to calibrate the formula in advance 
by performing a fully general 
relativistic simulation and by comparing the waveforms computed by the 
quadrupole formula with those computed from the metric. 
The quadrupole formula adopted in our study has been 
calibrated in simulations for highly relativistic, highly 
oscillating, and rapidly rotating neutron stars \cite{SS}. Thus, 
we believe that the quadrupole formula adopted in this paper 
is appropriate and that the numerical results presented here are 
approximate solutions of high quality. 

In this paper, we have focused on the neutron star formation and on 
the comparison with a previous work \cite{HD}. 
If more massive progenitor is chosen as the initial condition, 
a black hole instead of a neutron star will be formed. 
Formation of black holes and corresponding gravitational waves 
have been studied by several groups \cite{N,SP,shiba2000}. 
However, the initial conditions in their previous works are not 
very realistic for modeling a rotating stellar core collapse in nature.
Namely, the stellar core collapse to a black hole from a realistic
initial condition 
in fully general relativistic simulation is an unsolved issue. 
We are currently working in this subject and 
will present the numerical results in a subsequent paper. 

\vspace{4mm}

\begin{center}
{\bf Acknowledgments}
\end{center}

We thank Toni Font for discussion and careful reading of this manuscript,
and Harald Dimmelmeier for comments. 
We also thank Harald Dimmelmeier, Toni Font, 
Jose-Maria Ib\'a\~nez, and Eward M\"uller 
for suggesting this work. Numerical computations were performed 
on the FACOM VPP5000 machine in the data processing center of 
National Astronomical Observatory of Japan.
This work is in part supported by Japanese Monbu-Kagakusho Grants 
(Nos. 14047207, 15037204, and 15740142).


\begin{thebibliography}{99}

%%\bibitem{hyper} S. E. Woosley, Astrophys. J. {\bf 405}, 273 (1993): 
%%B. Paczynski, Astrophys. J. {\bf 494}, L45 (1998):  
%%A. I. MacFadyen and S. E. Woosley, Astrophys. J. {\bf 524}, 262 (1999): 
%%A. I. MacFadyen, S. E. Woosley and A. Heger, Astrophys. J. {\bf 550}, 410
%%(2001). 

\bibitem{Siebel} F. Siebel, J. A. Font, E. M\"uller, and P. Papadopoulos,
Phys. Rev. D {\bf 67}, 124018 (2003). 

\bibitem{Newton} L. S. Finn and C. R. Evans, Astrophys. J. {\bf 351}, 
588 (1990).

\bibitem{Newton1} R. M\"onchmeyer, G. Sch\"afer, E. M\"uller and R. 
Kates, Astron. and Astrophys. {\bf 246}, 417 (1991). 
E. M\"uller, M. Rampp, R. Buras, H.-T. Janka, and D. H. Shoemaker,
astro-ph/0309833. 
%%E. M\"uller and H.-T. Janka, Astron. Astrophys. {\bf 103}, 358 (1997).

\bibitem{Newton2}
S. Bonazzola and J.-A. Marck, Astron. Astrophys. {\bf 267}, 623 (1993). 

\bibitem{Newton3}
S. Yamada and K. Sato, Astrophys. J. {\bf 434}, 268 (1994):
{\bf 450}, 245 (1995): K. Kotake, S. Yamada, and K. Sato,
Phys. Rev. D {\bf 68}, 044023 (2003). 

\bibitem{Newton4} T. Zwerger and E. M\"uller, Astron. Astrophys.
{\bf 320}, 209 (1997).

\bibitem{Newton45}
M. Rampp, E. M\"ulelr and M. Ruffert, Astron. Astrophys.
{\bf 332}, 969 (1998). 

\bibitem{Newton5}
C. Fryer and A. Heger, Astrophys. J. {\bf 541}, 1033 (2000):
C. Fryer, D. E. Holz and A. Heger, Astrophys. J. {\bf 565}, 430 (2002). 

\bibitem{Newton6} C. D. Ott, A. Burrows, E. Livne, and R. Walder,
astro-ph/0307472. 

\bibitem{HD} H. Dimmelmeier, J. A. Font and E. M\"uller,
Astro. Astrophys. {\bf 388}, 917 (2002); {\bf 393}, 523 (2002). 

\bibitem{IWM} J. Isenberg and J. Nester, in {\sl General Relativity 
and Gravitation} Vol.1, edited by A. Held, (Plenum Press, New York 1980);
Waveless Approximation Theories of Gravity, preprint (1978), 
University of Maryland.

\bibitem{SS} M. Shibata and Y. Sekiguchi, Phys. Rev. D {\bf 68}, 104020
(2003). 

\bibitem{CST96} G. B. Cook, S. L. Shapiro, and S. A. Teukolsky,
Phys. Rev. D {\bf 53}, 5533 (1996). 

%%\bibitem{Siebel0} F. Siebel, J. A. Font, E. M\"uller and P. Papadopoulos,
%%Phys. Rev. D {\bf 65}, 064038 (2002). 

\bibitem{S2002} M. Shibata, Phys. Rev. D {\bf 67}, 024033 (2003). 

\bibitem{Font}
J. A. Font, Living Review Relativity {\bf 3}, 2, 2000
http://www.livingreviews.org/Articles/Volume2/2000-2font: F. Banyuls, 
J. A. Font, J.-Ma. Ib\'a\~nez, J. M. Marti, and J. A. Miralles, 
Astrophys. J. {\bf 476}, 221 (1997). 

\bibitem{shibata} M. Shibata, Prog. Theor. Phys. {\bf 101}, 1199 (1999); 
Phys. Rev. D {\bf 60}, 104052 (1999).

\bibitem{shiba2000} M. Shibata, 
Prog. Theor. Phys. {\bf 104}, 325 (2000).

\bibitem{SBS} M. Shibata, T. W. Baumgarte and 
S. L. Shapiro, Phys. Rev. D {\bf 61}, 044012 (2000); 
Astrophys. J. {\bf 542}, 453 (2000). 

\bibitem{bina} M. Shibata and K. Ury\=u, Phys. Rev. D {\bf 61}, 064001
(2000): Prog. Theor. Phys. {\bf 107}, 265 (2002);
M. Shibata, K. Taniguchi, and K. Ury\=u, Phys. Rev. D {\bf 68},
(2003). 

\bibitem{SN} M. Shibata and T. Nakamura, Phys. Rev. D {\bf 52}, 5428 (1995): 
In \cite{shibata,SBS,bina} and this paper,
we adopt a formulation slightly modified from the original 
version presented in this reference. 

%%\bibitem{NOK} T. Nakamura, K. Oohara, and Y. Kojima, 
%%Prog. Theor. Phys. Suppl. {\bf 90}, 1 (1987). 

\bibitem{S03} M. Shibata, Astrophys. J. {\bf 595}, 992 (2003). 

\bibitem{alcu} M. Alcubierre, S. Brandt, B. Br\"ugmann, 
D. Holz, E. Seidel, R. Takahashi and J. Thornburg,
Int. J. Mod. Phys. D {\bf 10}, 273 (2001). 

%\bibitem{moncrief} V. Moncrief, Ann. of Phys. {\bf 88}, 323 (1974).
%
%\bibitem{T82} S. A. Teukolsky, Phys. Rev. D {\bf 26}, 745 (1982). 
%
%\bibitem{CST} G. Cook, S. L. Shapiro, and S. A. Teukolsky, 
%Astrophys. J. {\bf 422}, 227 (1994). 

\bibitem{ST} For example, S. L. Shapiro and S. A. Teukolsky, {\em Black
Holes, White Dwarfs, and Neutron Stars} (Wiley Interscience, 
New York, 1983).

\bibitem{KEH} See, e.g., H. Komatsu, Y. Eriguchi, and 
I. Hachisu, Mon. Not. R. Astron. Soc., {\bf 237}, 355 (1989);
{\bf 239}, 153 (1989). 

\bibitem{Ster} See N. Stergioulas, Living Rev. Relativ, {\bf 1}, 8 (1998) 
for a historical review about computation of relativistic rotating stars. 

\bibitem{SKS} M. Shibata, S. Karino, and Y. Eriguchi,
Mon. Not. R. Astron. Soc., {\bf 334}, L27 (2002): {\bf 343}, 619 (2003). 

\bibitem{SS02} M. Shibata and S. L. Shapiro, Astrophys. J. Lett. {\bf 572},
L39 (2002). 

\bibitem{shiba03} M. Shibata, Astrophys. J. (2004), in press. 

\bibitem{Tas} J.-L. Tassoul, {\em Theory of Rotating Stars} 
(Princeton University Press, Princeton, New Jersey, 1978). 

\bibitem{Ch64} S. Chandrasekhar, Astrophys. J. {\bf 140}, 417 (1964). 

\bibitem{N} T. Nakamura, Prog. Theor. Phys. 
{\bf 65}, 1876 (1981); {\bf 70}, 1144 (1983).

\bibitem{SP} R. F. Stark and T. Piran, Phys. Rev. Lett. 
{\bf 55}, 891 (1985); in {\it 
Dynamical Spacetimes and Numerical Relativity}, 
edited by J. M. Centrella (Cambridge University Press,
Cambridge, England), p. 40.

\end{thebibliography}
\end{document}